\documentclass[aip,reprint]{revtex4-2}

\usepackage{graphicx}
\usepackage{dcolumn}
\usepackage{bm}
\usepackage{mathtools}
\usepackage{amssymb}
\usepackage{amsthm}
\usepackage{amsmath}
\usepackage{tikz}
\usepackage{tabularx}
\usepackage{listings}
\usepackage{changes}
\usepackage{xcolor}
\DeclarePairedDelimiterX\braket[2]{\langle}{\rangle}{#1\,\delimsize\vert\,\mathopen{}#2}

\begin{document}

\title{\texttt{TENSO}: Software Package for Numerically Exact Open Quantum Dynamics Based on Efficient Tree Tensor Network Decomposition of the Hierarchical Equations of Motion}

\author{Juan C. Rodriguez-Betancourt}
 \affiliation{Department of Chemistry, University of Rochester, Rochester, New York 14627, United States}
  \altaffiliation{These authors contributed equally}
  \author{Michelle C. Anderson}

 \affiliation{Department of Chemistry, University of Rochester, Rochester, New York 14627, United States}
  \altaffiliation{These authors contributed equally}
\author{Luchang Niu}
 \affiliation{Department of Physics and Astronomy, University of Rochester, Rochester, New York 14627, United States}
\author{Xinxian Chen}
 \affiliation{Department of Chemistry, University of Rochester, Rochester, New York 14627, United States}
\author{Ignacio Franco}
 \email{ignacio.franco@rochester.edu}
\affiliation{Department of Chemistry, University of Rochester, Rochester, New York 14627, United States}
\affiliation{Department of Physics and Astronomy, University of Rochester, Rochester, New York 14627, United States}
\affiliation{The Institute of Optics, University of Rochester, Rochester, New York 14627, United States}

\date{\today}

\begin{abstract}
\texttt{TENSO} is a versatile and powerful open-source software package for numerically exact simulations of the dynamics of quantum systems immersed in structured thermal environments. 
It is based on a tree tensor network decomposition of the hierarchical equations of motion (HEOM) that efficiently curbs its curse of dimensionality with bath complexity.
As such, \texttt{TENSO} enables exact non-Markovian open quantum dynamics simulations even with complex environments typical of chemistry and quantum information science. \texttt{TENSO} allows for time-dependent drive in the system, and for non-commuting fluctuations. More generally, \texttt{TENSO} efficiently propagates the dynamics for any method with a generator of the dynamics that can be expressed in a sum-of-products form, including the HEOM and multi-layer multiconfigurational time-dependent Hartree methods. \texttt{TENSO} enables simulations using tensor trees and trains of arbitrary order, and implements three propagation strategies for the coupled master equations; two fixed rank methods that require a constant memory footprint during the dynamics and one adaptive rank method with a variable memory footprint controlled by the target level of computational error. In contrast to the accompanying theory and algorithmic paper [J. Chem. Phys. \textbf{163}, 104109 (2025)] the focus here is on the practical usage and applications of \texttt{TENSO} with underlying theoretical concepts introduced only as needed. 
\end{abstract}

\maketitle

\section{\label{sec:level1}Introduction}

Open quantum systems refer to systems interacting with their quantum environment. This interaction introduces quantum noise or decoherence and is responsible for ubiquitous processes in nature such as dephasing, energy relaxation and the thermalization of quantum systems making a detailed understanding of open quantum dynamics essential in chemistry, physics, and quantum information science.\cite{breuer_theory_2007, nielsen2010,weiss_quantum_2021, schlosshauer2007, rivas_open_2012, weimer2021, xu2026} Further, elucidating decoherence mechanisms is necessary to engineer quantum environments for quantum and chemical control tasks.\cite{zhan2026, siddiqi_engineering_2021,gustin_mapping_2023,kim2022,kim2024,gustin2026} The aim of \texttt{TENSO}, Tensor Equations for Non-Markovian Structured Open Systems, \cite{chen2025,TENSO} is to predict the dynamics of quantum systems interacting with macroscopic thermal environments of arbitrary complexity using the numerically exact hierarchical equations of motion (HEOM).\cite{tanimura_numerically_2020} \texttt{TENSO} combines the bexcitonic generalization of the HEOM~\cite{chen_bexcitonics_2024} --which unites HEOM variants and exposes their mathematical structure-- with a tree tensor network (TTN) decomposition which enables numerically efficient simulation.

In open quantum dynamics it is customary to decompose the Hamiltonian, $H=H_\textrm{S} + H_\textrm{SB} + H_\textrm{B}$, of the quantum universe into Hamiltonians of a system $H_\textrm{S}$, a bath $H_\textrm{B}$ and their interaction $H_\textrm{SB}$.\cite{breuer_theory_2007} Since the bath is usually macroscopic, the strategy in open quantum dynamics is to develop quantum master equations (QMEs) that capture its influence on the system without propagating the bath degrees of freedom explicitly. A variety of approaches have been developed to simulate open quantum system dynamics.\cite{vega2017,breuer_theory_2007,weimer2021,nitzan2024, xu2026, Mulvihill2021,landi2024,anto2023, ankerholdrmp2026} Of particular interest are numerically exact QMEs as they can be used to model a large class of problems of importance in chemistry and quantum information science with an accuracy that can be assured. This contrasts with common strategies based on the Born-Markov approximation that only apply to systems weakly coupled to fast environments with short memory time,\cite{breuer_theory_2007} conditions that are often too restrictive for chemical dynamics.

The HEOM is a powerful method for numerically exact open quantum dynamics for systems embedded in thermal Bosonic environments.\cite{tanimura_numerically_2020,ikeda_generalization_2020,ishizaki2006,tanaka2010, chen_bexcitonics_2024} In the case of linear system-bath coupling, the bath correlation function (BCF) contains all information about the bath needed to construct the quantum master equation. To make the QME numerically tractable, HEOM decomposes the BCF into a series of $K$ complex decaying exponentials which allows the influence of the environment to be captured by a hierarchy of auxiliary density matrices (ADMs). The HEOM has been employed effectively for models of spectroscopy, and for charge and energy transport in photosynthetic complexes, molecular aggregates, Holstein models, and organic photovoltaic cells.\cite{bai2021,kramer2018,seibt2021,yan2018,cainelli2021,yan2020,jankovic2023} The main challenge of HEOM is that the cost of the quantum dynamics grows exponentially with the number of features $K$ in the bath, thus limiting the use of HEOM to simple model problems with unstructured environments.

Due to its importance, several computational packages implement HEOM simulations, including \textsc{QuTiP}, \textsc{pyHEOM}, \textsc{HierarchicalEOM.jl}, \textsc{GPU-HEOM}, and \textsc{DM-HEOM}.\cite{lambert_qutip_2025,ikeda_generalization_2020,huang_efficient_2023,kreisbeck_long-lived_2012,kramer2018} 
However, these implementations are often limited by the number of features $K$ that can be included, and present-day capabilities are insufficient to capture chemically realistic environments, especially when low temperature correction terms are required.

To mitigate this curse of dimensionality, recently we introduced a TTN decomposition of the HEOM (TTN-HEOM) which allows for efficient compression of the simulation space thus enabling simulations of open quantum systems interacting with environments with large numbers of features.\cite{chen2025} Our approach further allows for time-dependent system Hamiltonians and multiple baths, providing a useful and versatile method to investigate dissipative dynamics. 

Tensor network decompositions represent the state of the art in the simulation of many-body systems.\cite{schollwock2011,wang2003,lyu2022,tamascelli2019,bose2022,cygorek2022,kundu2023, strathearn2018} In particular, they are the basis for the powerful multi-layer multi-configurational time-dependent Hartree (ML-MCTDH)\cite{wang2003,meyer1990, Lindoy2025} method which is the gold standard in high-dimensional closed quantum dynamics simulations. Our TTN-HEOM developments are parallel to the ML-MCTDH as the three main principles (sum of product dynamical generator, TTN decomposition, and Dirac-Frenkel time-dependent variational principle (TDVP)\cite{meyer1990,raab2000,murnaghan1935,dirac1930} to isolate the equations of motion) are identical. However, while ML-MCTDH is designed for unitary dynamics, the TTN-HEOM is designed for thermal dissipative dynamics.

\texttt{TENSO} employs tensor network approaches by arranging the large set of ADMs in the HEOM formalism into an extended density operator (EDO). This higher order EDO is decomposed into core tensors using the TTN representation. To optimally evolve the TTN such that it correctly represents the EDO, \texttt{TENSO} applies the TDVP to develop QMEs for the core tensors. The \texttt{TENSO}  implementation of this QME provides efficient, stable, error-controlled evolution best suited for open quantum dynamics with strong coupling and highly structured bath spectral densities. 

We illustrate the usage of \texttt{TENSO} and demonstrate its versatility and wide-applicability to open quantum dynamics through three examples. First is the quintessential spin-boson problem which is a key model in spectroscopy and quantum information.\cite{Leggett1987,mukamel1995principles, nitzan2024}
Using this model, we will illustrate the fundamental features of \texttt{TENSO}, how to perform simulations with structured spectral densities, how \texttt{TENSO} easily addresses non-commuting fluctuations, how to invoke time-dependent Hamiltonians, and how to manipulate the convergence. We will also briefly mention \texttt{TENSO}'s extension to MCTDH, emphasizing the adaptability of \texttt{TENSO}'s underlying structure to other master equations with propagators in sum-of-products form. We then address the Fenna-Matthews-Olson (FMO) \cite{fenna1975, fenna1979} complex, a paradigmatic multilevel model problem in the study of photosynthetic energy capture.\cite{chenu2015,duan2022,hu2022,cao2020,jang2018, gustin2025} 
We use it to demonstrate how to employ \texttt{TENSO} to address multilevel systems coupled to multiple structured baths. Our third example is the sudden death of entanglement between two qubits interacting with a thermal environment.\cite{yu2009}
We use it to demonstrate how \texttt{TENSO} can address problems in quantum information. 

It is useful to frame \texttt{TENSO} within the increasing ecosystem of tensor-network based simulation methods. The packages pyTTN\cite{Lindoy2025} and QuTree\cite{ellerbrock2024} provide frameworks for tensor operations. TTNO\cite{li_optimal_2024} yields optimized sum-of-products decomposition of general operators. RENORMALIZER\cite{renormalizer} and MPSDynamics.jl\cite{mpsdynamicsjl_2024} provide tensor-based time-dependent density matrix renormalization group (TD-DMRG) and time evolving density operator with orthogonal polynomials (TEDOPA) computations.\cite{chen2026} The Heidelberg MCTDH package focuses on wavefunction based dynamics.\cite{heidelberg} In contrast, TENSO is focused on HEOM calculations. The mpsqd,\cite{guan2024} also implements tensor-based HEOM but using tensor trains instead of tensor trees. Together TENSO’s HEOM focus, accessible python implementation, and ability to address general tree structures with both adaptive and fixed rank propagation strategies differentiate it from other options and make it a versatile tool in this ecosystem.

The structure of this paper is as follows. Section~\ref{sec:theory} reviews the bexcitonic HEOM and its TTN decomposition. Section~\ref{sec:numexamples} demonstrates the installation, usage and capabilities of \texttt{TENSO} through representative numerical examples. Section \ref{sec:defaults} summarizes typical simulation parameters appropriate for most problems, and section \ref{sec:scaling} computational cost. Section~\ref{sec:software} provides information on the architecture and design of the \texttt{TENSO} package, and section \ref{sec:advanced} discusses modifications which can be made to the \texttt{TENSO} code to address specialized or more advanced applications.

\section{Methods}

We now provide an overview of the most important theory underlying \texttt{TENSO}. Refs. \onlinecite{chen_bexcitonics_2024} and \onlinecite{chen2025} provide detailed derivations, and a full discussion of the propagation strategies employed. Throughout we use atomic units where $\hbar=1$.
\label{sec:theory}
\subsection{Bath Correlation Function and its Decomposition into Features}

The HEOM focuses on the open quantum dynamics of systems immersed into an environment that can be described as a collection of harmonic oscillators with Hamiltonian
\begin{equation}
    {H}_\textrm{B} = \sum_j \left( \frac{p_j^2}{2 m_j} + \frac{m_j \omega_j^2 x_j^2}{2} \right).
\end{equation}
Here, $x_j$ and $p_j$ are the position and momentum of the $j$-th oscillator with frequency $\omega_j$ and effective mass $m_j$. The system-bath coupling Hamiltonian is of the form $ H_\textrm{SB} =  Q_\textrm{S} \otimes  X_\textrm{B}$ where $ Q_\textrm{S}$ is a system operator and $ X_\textrm{B} = \sum_j c_j x_j$ is a collective bath coordinate. Here, $c_j$ quantifies  the coupling strength between the system operator and the $j$th bath mode. Harmonic models of the environment are broadly used as any environment can be represented by this model up to second order in perturbation theory.\cite{feynman1963,caldeira1983} Further, this is commonly applicable in the thermodynamic limit as the system-bath interaction is diluted over a macroscopic number of modes.\cite{suarez1991,makri1999}
For harmonic environments, the dynamical properties of the bath are completely captured by the BCF, $C(t) = \langle  \tilde{X}_\textrm{B}(t)  \tilde{X}_\textrm{B} (0)\rangle$, where $\tilde{X}_\textrm{B}(t)$ is the collective bath coordinate in the interaction picture of $H_\textrm{S}+ H_\textrm{B}$ and the expectation value is over the bath thermal state.\cite{nitzan2024} 
Formally the BCF can be expressed as
\begin{equation}
    C(t) = \int_{-\infty}^{+\infty} \mathcal{J}(\omega) f_{\text{BE}}(\beta \omega) e^{-i \omega t} \, \mathrm{d}\omega,
\end{equation}
where 
\begin{equation}
\mathcal{J}(\omega) = \sum_j\frac{ |c_j|^2}{ (2m_j \omega_j)} \left[ \delta(\omega - \omega_j) - \delta(\omega + \omega_j) \right]
\end{equation}
is an odd extension of the bath spectral density, a quantity that summarizes the frequencies $\{\omega_j\}$ of the oscillators in the bath and specifies their coupling strengths to the system. The function $f_{\text{BE}}(\beta \omega) = \left(1 - e^{-\beta \omega} \right)^{-1}
$ is related to the Bose--Einstein distribution, and $\beta=1/k_BT$ is the inverse temperature. Without loss of generality, the residue theorem is used to decompose the BCF into a sum of complex exponential functions referred to as features, typically via Padé\cite{hu2010} or Matsubara\cite{Ishizaki2005} expansions. This allows us to write the BCF as
\begin{equation}\label{eq:bcfdecores}
\begin{split}
&C(t) ={}  -2\pi i \sum_i \operatorname{Res}_{z = \zeta_i} \left[ \mathcal{J}(z) \right] f_{\text{BE}}(\beta \zeta_i) e^{-i \zeta_i t} \\
& - 2\pi i \sum_j \operatorname{Res}_{z = \xi_j} \left[ f_{\text{BE}}(z) \right] \mathcal{J}(\xi_j / \beta) e^{-i (\xi_j / \beta) t},
\end{split}
\end{equation}
where $\left \{ \zeta_i  \right \} $ are the first-order poles in the lower half of the complex plane of $\mathcal{J}(\omega)$ and $\left \{ \xi_i  \right \} $ are those of $f_{\text{BE}}(\beta \omega)$. By this method, the BCF can always be decomposed in terms of complex decaying exponentials such that $C(t) = \sum_{k=1}^K c_k e^{\gamma_k t}$ and $C^*(t) = \sum_{k=1}^K \bar{c}_k e^{\gamma_kt}$ where $c_k, \bar{c}_k$ and $\gamma_k$ are complex numbers. Each term $k$ in Eq.~(\ref{eq:bcfdecores}) corresponds to a feature and $K$ is the total number of features.\cite{chen_bexcitonics_2024} The first set of features arises due to the spectral density itself, while the second set arises due to the thermal factor and is referred to as low temperature corrections.

Two common models for the spectral density are the Drude-Lorentz (DL) model and the Brownian oscillator (BO) model.\cite{nitzan2024,mukamel1995principles} The DL spectral density,
\begin{equation}\label{eq:DL}
    J_\textrm{DL}(\omega) = \frac{2\lambda}{\pi} \frac{\omega_c \omega}{\omega^2 + \omega_c^2}
\end{equation}
describes an Ohmic environment with $\lambda$ reorganization energy and $\omega_c$ cut-off frequency. Drude-Lorentz environments contribute to a single, simple decaying exponential with characteristic relaxation timescale of $\omega_c^{-1}$.
In turn, the BO spectral density is expressed as
 \begin{equation}\label{eq:BO}
     J_\textrm{BO}(\omega) = \frac{4\lambda}{\pi} \frac{\eta \omega_0^2 \omega}{\left( \omega^2 - \omega_0^2 \right)^2 + 4 \eta^2 \omega^2},
\end{equation}
and describes the contribution of a discrete harmonic oscillator where $\lambda$ is its reorganization energy, $\omega_0$ its natural frequency and $1/\eta$ its damping rate. Brownian oscillators contribute two oscillatory-decaying features to the BCF decomposition. In \texttt{TENSO}, the actual spectral density used for the BO oscillator is
\begin{equation}
    J_{\mathrm{BO}}(\omega) = \frac{4 \lambda \eta }{\pi} \frac{(\omega_0'^2 + \eta^2)\omega }{[(\omega + \omega_0')^2 + \eta^2][(\omega -\omega_0')^2 + \eta^2]}
\end{equation}
which is mathematically equivalent to the standard Brownian oscillator in Eq.\eqref{eq:BO} under the parameter mapping $\omega_0' = \sqrt{\omega_0^2 - \eta^2}$, which corresponds to the effective frequency under damping ($\eta < \omega_0$). The factored form is used because its relevant poles are simply located at $\pm \omega_0' - i\eta$, which makes the decomposition in Eq. \eqref{eq:bcfdecores} straightforward. In chemical dynamics it is often possible to define a spectral density as a combination of DL and BO terms as\cite{gustin_mapping_2023}
\begin{equation}
    J(\omega)=J_\textrm{DL}(\omega)+\sum_\textrm{b}J_\textrm{BO}^{\textrm{(b)}}(\omega).
\end{equation}

Environments with a large number of features, $K$, are referred to as highly structured. Since the numerical cost of the open quantum dynamics scales exponentially with $K$, the computations become increasingly challenging. In fact, HEOM calculations using standard methods are only feasible with one to five features. By contrast, in \texttt{TENSO} the spectral density which describes the thermal environment can be highly structured as needed to describe baths of chemical complexity. However, the environment itself must be harmonic or, alternatively, mapped to a surrogate harmonic bath, which has been shown to serve as an accurate model of condensed phase environments.\cite{cho2025}

\subsection{Hierarchical Equations of Motion}

While the full dynamics of the composite system with density matrix $\rho(t)$ is unitary, the dynamics of the reduced system, ${\rho}_\textrm{S} = \mathrm{Tr_B} (\rho(t))$ is non-unitary, where $\mathrm{Tr_B}$ indicates a trace over the bath degrees of freedom. The dynamical map of $\rho_\mathrm{S}(t)$ is,\cite{Tanimura1989}
\begin{equation}
    \tilde\rho_\textrm{S} (t) = \mathcal{T} \tilde{\mathcal{F}}(t,0)\rho_\textrm{S}(0)
\end{equation}
with time-ordering operator $\mathcal{T}$ and
\begin{equation}
    \tilde{\mathcal{F}}(t,0) = e^{-\int_0^t ds \tilde Q_\textrm{S}^\times(s) \int_0^s du \left(C(s-u) \tilde{Q}_\textrm{S}(u) \right)^\times}.
\end{equation}
Throughout, the tilde indicates operators  in the interaction picture of  $H_\textrm{0} (t) = H_\textrm{S}(t) + H_\textrm{B}$,
\begin{equation}
    \tilde{O}(t) = (\mathcal{T} e^{-i \int_0^t H_0 (t') dt'})^\dagger O(t) (\mathcal{T} e^{-i \int_0^t H_0 (t') dt'}).
\end{equation}
For any operator $A$ we use the convention $A^\times B= A^> B- A^< B=  AB - BA^\dagger$. 

Employing the decomposition of the structured bath into features following Eq. (\ref{eq:bcfdecores}) and calculating the time-derivatives in the formal evolution of $\tilde{\rho}_\textrm{S}(t)$ reveals that the influence of the thermal environment is exactly captured by an infinite hierarchy of auxiliary density matrices (ADMs), $\{\varrho_{\vec{n}}(t)\}$, of the same dimension, $M\times M$, as the system density matrix. Here $\vec{n}=(n_1,...n_k,...n_K)$ is a $K$ dimensional index, with $K$ being the number of features of the bath. Each $n_k$ runs formally from $0$ to infinity and practically from $0$ to a truncation limit known as the hierarchy depth. 

\texttt{TENSO} takes the generalized bexcitonic view of HEOM\cite{chen_bexcitonics_2024} where the system is conceptualized as interacting with a collection of fictitious, bosonic quasiparticles called bexcitons, each corresponding to a bath feature. In the bexcitonic HEOM, the ADMs are arranged into an extended density operator (EDO). The EDO is $|\Omega(t)\rangle = \sum_{\vec{n}} \varrho_{\vec{n}} (t) |\vec{n}\rangle$ which is specified in the basis given by $|\vec{n} \rangle = |n_1\rangle \otimes ...\otimes|n_k\rangle  \otimes ... \otimes |n_K\rangle$ such that a given ADM is found as $\varrho_{\vec{n}}(t) = \langle \vec{n}| \Omega(t)\rangle$. The system density matrix is $\rho_\textrm{S}(t) =\varrho_{\vec{0}}(t) = \langle \vec{0}|\Omega(t)\rangle$, meaning it is found where all indices in $|\vec{n}\rangle$ are zero.

The equation of motion of the EDO is given by,\cite{chen_bexcitonics_2024,chen2025}

\begin{equation}\label{eq:heom-std}
    \frac{\rm d}{\rm dt}  |\Omega(t)\rangle =
   \left(-iH_\textrm{S}^\times (t) + \sum_{k=1}^K \mathcal{D}_k   \right)|\Omega(t)\rangle,
\end{equation}
 where $\mathcal{D}_k$ is the dissipator associated with the $k$-th bath feature, or bexciton. In particular, for each $k$,
\begin{equation}\label{eq:dissipator}
    \mathcal{D}_k = \gamma_k \hat \alpha_k ^\dagger \hat \alpha_k + \left(c_k Q_S^> - \bar{c}_k Q_\textrm{S}^<  \right) \hat{z}_k^{-1} \hat \alpha_k^\dagger - Q_\textrm{S}^\times \hat \alpha_k \hat z_k
\end{equation}
with $\hat{z}_k$, the metric, being any invertible operator which commutes with $\hat \alpha_k^\dagger \alpha_k$, $\hat{\alpha}_k^\dagger$ and $\hat{\alpha}_k$ are the bosonic creation and annihilation operators for the $k$th bexciton, $([ \hat \alpha_k , \hat\alpha_{k'}^\dagger]) = \delta_{k,k'}$. In the bexcitonic view of HEOM, each feature, $k$, is associated with a fictitious, bosonic quasiparticle. The operator $\hat \alpha^\dagger_k$ creates a $k$th bexciton and the operator $\hat \alpha_k$ destroys it. With $|\vec{n}\rangle$ we associate a state with $|n_k\rangle$ bexcitons created for each $k$. The $K$ bexcitons constitute a coarse grained, but numerically exact, representation of the macroscopic bath but are not physical particles associated with modes of the bath. Although useful for monitoring convergence of the dynamics, bexcitons do not provide physical information about the bath state. 

For each feature $k$, the bexcitonic ladder must be truncated at a maximum occupation $(N_k-1)$, the depth of the $k$th-bexciton, to allow for practical simulations.\cite{chen_bexcitonics_2024}
For an $M$-state system coupled to $K$ bath modes, each truncated at $N_k=\mathcal{O}(N)$, the memory cost of the EDO scales as $\mathcal{O}(M^2 N^K)$. This exponential growth of memory requirements with $K$ constitutes the principal limitation of HEOM simulations.\cite{chen2025}  This problem is especially severe for highly structured spectral densities and at low temperatures where many features are required. 

Equation (\ref{eq:heom-std}) specifies a class of QMEs. Conventional HEOM uses $\hat{z}_k = i(\hat{\alpha}_k^\dagger \hat{\alpha}_k)^{-1/2}$ and a number basis representation of $|\vec{n}\rangle$. By adjustment of the metric and basis representation of $|\vec{n}\rangle$, we can recover and develop many variants of HEOM.\cite{chen_bexcitonics_2024}

\subsection{Tree Tensor Network Decomposition}
\label{sec:tn}
The curse of dimensionality in HEOM can be mitigated by recognizing that the EDO contains significant redundancy that can be systematically compressed via tensor network representations, greatly reducing memory requirements.\cite{chen2025, Ke2023,ke2022,yan2021}  We begin by reexpressing the EDO. Analogously to the density matrix of the system, $\rho_\mathrm{S}(t)$, with matrix elements $[\rho_\mathrm{S}]_{ij}=\langle i|\rho_\mathrm{S}(t)|j\rangle$ in a basis $\{\lvert i\rangle\}$ spanning the system Hilbert space, the EDO can be represented as
\begin{equation}
  [\Omega(t)]_{i j n_1 \cdots n_K} 
  = \langle i \,| \,\langle n_1 \cdots n_K \,|\, \Omega(t)\rangle \,|\, j \rangle,
  \label{eq:EDO-elements}
\end{equation}
where $\{\lvert n_k\rangle\}_{n_k=0}^{N_k-1}$ is the truncated number basis for the $k$th-bexciton. The dynamics of $\Omega(t)$ in Eq. (\ref{eq:heom-std}) can be written compactly as\cite{chen2025}
\begin{equation}
  \frac{\rm d}{\rm dt}\,\Omega(t) = \mathcal{L}(t)\,\Omega(t),
  \label{eq:EDO-eom}
\end{equation}
where $\mathcal{L}(t)$ is the tensor representation of the Liouvillian superoperator generating the dynamics in Eq. (\ref{eq:heom-std}). Because the basis factorizes as $\lvert i\rangle\otimes \langle j\rvert\otimes \lvert n_1\rangle\otimes\cdots\otimes \lvert n_K\rangle$, $\mathcal{L}(t)$ admits the operator in sum-of-products (SoPs) form,
\begin{equation}
\label{eq:master_lvn}
\mathcal{L}(t) \equiv\sum_{m=1}^{5 K+2} h_m^{>}(t) \otimes h_m^{<}(t) \otimes h_m^{(1)} \otimes \cdots \otimes h_m^{(K)} 
\end{equation}
with local operators $h_m^{(\kappa)}$ where the index $m$ runs over all terms in Eq. (\ref{eq:heom-std}) and $(\kappa = >, <, 1, \ldots,K)$. Each dissipator $\mathcal{D}_k$ contributes five terms and the unitary Liouvillian, $-iH_S^{\times}(t)$, contributes two. Here $h_m^{>}(t)$ and $h_m^{<}(t)$ act on the system space while $h_m^{(k)}$ acts on the truncated $k$th-bexciton space.

The EDO is a high dimensional tensor. The memory requirements to describe a tensor are exponential in the tensors order, meaning that for a tensor $A_{a_1...a_D}$, where the $a_i = 1,...,R$, the tensor requires $O(R^D)$ memory space. Performing mathematical operations on the tensor which require access to all elements thus also scales catastrophically. To avoid this, the EDO is compressed by expressing it as a network of low-order core tensors. The tree is assembled in general by repeated applications of the singular value decomposition\cite{eckart1936} (SVD), a procedure known as the hierarchical Tucker decomposition (HTD).\cite{grasedyck2010,grasedyck2011} When performing the decomposition, small singular values judged to be relatively unimportant  discarded, with the number of retained singular values known as the rank of the bond, $R$. Contractions are performed between the network's core tensors according to the topology of a tree graph and the lowest order tensors possible are employed for efficiency. The lowest order that allows for the assembly of an arbitrary size TTN is three. This decomposition $K$ order-three core tensors with $K-1$ contractions. 

The simplest TTN topology is a tensor train (TT),
\begin{equation}
\label{eq:omega_tensor}
\begin{aligned}
& \Omega_{i j n_1 \cdots n_K}= \\
& \quad \sum_{a_1 a_2 \cdots a_{K-1}}^{R_1 R_2 \cdots R_{K-1}} A_{i j a_1}^{(0)} U_{a_1 n_1 a_2}^{(1)} U_{a_2 n_2 a_3}^{(2)} \cdots U_{a_{K-1} n_{K-1} n_K}^{(K-1)} 
\end{aligned}
\end{equation}
where $\{a_s\}_{s=1}^{K-1}$ are bond indices of ranks $\{R_s\}$.   Similarly, a general TTN is written as,
\begin{equation}
\begin{aligned}
& {[\Omega(t)]_{i j n_1 \cdots n_K}=} \\
& \quad \sum_{a_1 \cdots a_{K-1}}^{R_1 \cdots R_{K-1}} A_{i j a_1}^{(0)} U_{a_1 \beta_1 \gamma_1}^{(1)} \cdots U_{a_{K-1} \beta_{K-1} \gamma_{K-1}}^{(K-1)} \\
& \quad \equiv\left[\operatorname{Con}\left(A^{(0)}(t), U^{(1)}(t), \ldots, U^{(K-1)}(t)\right)\right]_{i j n_1 \cdots n_K}
\end{aligned}
\end{equation}
where $\mathrm{Con}(\cdot)$ denotes contraction of the TTN core tensors, $A^{(0)}(t), U^{(1)}(t),...U^{(K-1)}(t)$, according to the chosen network topology, and each pair $(\beta_s,\gamma_s)$ corresponds either to a bath index $n_k$ (open bond) or to a virtual bond $a_u$ (internal contraction).

The semi-unitary core tensors $U^{(s)}(t)$ in the TTN are chosen to satisfy
\begin{equation}
  \sum_{\beta\gamma}
  \big[U^{(s)}(t)\big]^{\!*}_{a'_s, \beta \gamma}\;
  \big[U^{(s)}(t)\big]_{a_s, \beta \gamma}
  = \delta_{a'_s a_s},
  \label{eq:semiunitary}
\end{equation}
but the root tensor $A^{(0)}(t)$ is not constrained to be semi-unitary. Following standard practice, the system indices $(i,j)$ and the first bond $a_1$ are placed in the root tensor so that the physical system degrees of freedom remain uncompressed while the bath degrees of freedom are compacted.

For a hierarchy of depth $N$ and TTN bond ranks $R_s=\mathcal{O}(R)$, the TTN storage scales as
\begin{equation}
  \mathcal{O}\!\big(M^2 R \;+\; K N R (N+R)\big),
\end{equation}
thereby eliminating the exponential dependence on $K$ and providing a controllable, systematically improvable approximation that becomes exact as the ranks increase. The equation of motion for the tensor network is determined by the TDVP applied using the Liouvillian Eq. (\ref{eq:EDO-eom}). The derivation is detailed in Ref.~\onlinecite{chen2025} with the equations of motion for order three tensors in Eqs.~19 and 20, and their generalization to a TTN containing tensors with arbitrary order in the supplementary information in Eqs.~S14 and S15. In practice, the bond dimension or rank $R$ needs to be increased until convergence. The smaller the rank that can be used, the more efficient the compression of the TTN. To produce numerically exact results, \texttt{TENSO} also requires a converged hierarchy depth, low integration error, and sufficiently accurate decomposition of the BCF, as needed by any HEOM-based computation.

The most important result is the determination of equations of motion for the root and core tensors, admitting general numerical methods for solving coupled differential equations, for any master equation which admits a sum-of-products form for $\mathcal{L}(t)$. This allows flexibility in the implementation of propagation methods, including both fixed rank and adaptive rank integration schemes. 

\texttt{TENSO} implements three different propagation strategies, direct and two versions of projector splitting. The direct integration strategy, known as vmf in the package, is a fixed rank method that simultaneously integrates the non-linear coupled series of ordinary differential equations representing the evolution of the core and root tensors. These equations of motion involve inversion of matrices that are singular during the early steps of the dynamics, leading to the need for regularization techniques and, hence, regularization errors. This propagation strategy and the high-order regularization technique implemented in \texttt{TENSO} is detailed in Sec: IID1 in Ref.~\onlinecite{chen2025}. The advantage of this strategy is that it enables coupling the TTN-HEOM to well developed solvers of ordinary differential equations based on Runge-Kutta and other schemes that allow for larger integration time steps, adaptive time steps and even parallelization.

By contrast the projector splitting methods propagate the tensors in the network individually and sequentially based on Trotterization of the propagator. This approach avoids the regularization but is limited by the Trotterization errors which require small time steps. At each step of the projector-splitting methods, the root tensor is moved through a complete traversal of the tree, passing to an adjacent semi-unitary tensor via SVD at each sub-step. In this way, every core tensor is propagated while it is the root, for which the dynamics does not have the singularity issue. The two algorithms, ps1 and ps2, differ in that ps1 moves tensors to the root during propagation sweeps without adjusting their rank. In turn, ps2 merges neighboring tensors as it moves them to the root and performs an SVD to split the combined tensor back into two resulting in an adaptive rank algorithm. For detailed algorithms that are implemented in \texttt{TENSO}, see section II D in Ref.~\onlinecite{chen2025}.

\section{Numerical examples}\label{sec:numexamples}

\subsection{Obtaining and Installing \texttt{TENSO}}
\label{sec:examples}
\texttt{TENSO} is a python package which can be obtained from github:  \url{https://github.com/ifgroup/pytenso}. Users must have python $3.13$ or later installed to run \texttt{TENSO}. \texttt{TENSO} further relies on the packages \texttt{numpy},\cite{harris2020} \texttt{scipy},\cite{scipy} \texttt{pytorch},\cite{ansel2024} \texttt{torchdiffeq}\cite{torchdiffeq} and \texttt{tqdm}.\cite{tqdm} These packages provide high-level interfaces designed to simplify tensor operations across diverse computational platforms, including various CPU and GPU architectures. For this tutorial, users should also have \texttt{matplotlib}\cite{hunter2007} available. A python virtual environment should be prepared for \texttt{TENSO} and these packages made available in it prior to installation. A thorough guide to preparing virtual environments and installing python packages for Windows, Macintosh, and Linux users can be found in the official Python 3 documentation. Package managers Anaconda and Pip can also be used. Once the environment has been prepared and \texttt{TENSO} installed, run it as \lstinline{python <input file>}. \texttt{TENSO} reads all parameters of the system and bath from the input file using energy units of $\mathrm{cm^{-1}}$, time units of $\mathrm{fs}$, and temperature units of $\mathrm{K}$.

\subsection{Overview}

In this section we will use three sample systems, the spin-boson, the FMO complex, and a pair of qubits, all selected due to their important roles in quantum dynamics, to demonstrate how to employ \texttt{TENSO} on research problems ranging from energy transport to entanglement dynamics.

There are four main steps to running a TENSO calculation. First, libraries and modules are imported. Second, the bath correlation function decomposition is generated by the function \lstinline{gen_bcf}, or provided manually by an advanced user. Third, the initial system state, system-bath coupling, tensor network structure and propagator are initialized by the function \lstinline{system_multibath}. Finally, propagation is carried out to the desired final time using the tqdm package to display progress and estimate time remaining.

All examples in this tutorial are straightforward to run on a laptop or desktop computer with the exception of the Fenna-Matthews-Olson complex example which is computationally more demanding. \texttt{TENSO} excels at and provides the largest performance gains when handling difficult problems for HEOM, those which have complicated spectral densities which cannot be represented by a single Drude-Lorentz or Brownian bath alone, or those which need large hierarchy depths. 

\subsection{Example 1: Spin-Boson}
\label{sec:sbp1}
The spin-boson model, a single two level system interacting with a thermal boson bath,\cite{Leggett1987, nitzan2024} has been used to test the performance of quantum simulation methods, Markovian and non-Markovian,\cite{kumar2025,suarez2024,wenderoth2021,nesi2007} to study the impacts of bath parameters in open systems,\cite{deng2016,sun2025} and to understand physical processes involving two-level systems (TLS), ranging from spontaneous emission\cite{Leggett1987,peropadre2013} to superconducting qubit relaxation.\cite{kjaergaard2019,chirolli2007} It remains an important system of study.

The spin-boson system Hamiltonian is given by
\begin{equation}
    {H}_\textrm{S} = \frac{\Delta\epsilon}{2} {\sigma}_z + V {\sigma}_x,
\end{equation}
with energy gap $\Delta\epsilon$, tunneling amplitude $V$, and Pauli operators $ \sigma_z$ and $ \sigma_x$. The system-bath coupling is
\begin{equation}
    {H}_\textrm{SB} =  \frac{\sigma_z}{2} \otimes {X}_\textrm{B}.
\end{equation}
We will use the spin-boson problem to demonstrate the basic usage of \texttt{TENSO}, how to specify the spectral density of a structured bath, how to invoke a time-dependent Hamiltonian, and how to understand parameters that may impact numerical convergence of the results. 

\subsubsection{Dynamics in a structured bath}

Listing \ref{lst:full_file_example}, shows a complete \texttt{TENSO} input file specifying an HEOM calculation for a spin in interaction with a bath whose spectral density includes a DL and a BO term. The file is broken into four sections, (1) importing modules (2) bath correlation function definition (3) propagator setup and (4) propagation. All HEOM examples in this tutorial will require the same import statements. Simulation parameters used in Listing \ref{lst:full_file_example} are given in Table~\ref{tab:1}.

In \texttt{TENSO} the \lstinline{gen_bcf} helper function builds the BCF. It takes the cutoff frequencies, the reorganization energies, widths, and the resonant frequencies as lists that encode the spectral densities' structures. To label the two types of SDs, \texttt{TENSO} uses the suffixes \lstinline{_d} and \lstinline{_b} for the DL and Brownian SD, respectively. To use the Matsubara decomposition scheme, the argument \lstinline{'Pade'} is replaced with \lstinline{'Matsubara'} in the call to \lstinline{gen_bcf}. It is also possible to implement custom BCFS's and their decompositions, as detailed in section \ref{sec:advanced}.

Construction of the propagator in \texttt{TENSO} is performed with the \texttt{system\_multibath} function. The bath correlation object produced by \lstinline{gen_bcf} is needed by \lstinline{system_multibath} to construct the propagator.  The \lstinline{system_multibath} helper function can address one or more baths. It takes all system information, the BCF or BCF's in list form, system-bath coupling operators, and all additional simulation parameters such as integrator tolerances or the tensor network decomposition scheme. The tqdm package then uses the propagator to advance the system through time and displays a status bar of the calculation’s progress.

\begin{table}[htb]
\caption{\textbf{Simulation parameters for the spin--boson model.} The system parameters include the energy bias $\Delta\epsilon$ and the tunneling coupling $V$. The bath spectral density is modeled as a sum of Drude--Lorentz (DL) and Brownian oscillator (BO) contributions, parameterized by the reorganization energy $\lambda$, the cutoff frequency $\omega_c$, the damping coefficient $\eta$, and the vibrational frequency $\omega_0'$. The resulting spectral density is shown in Fig.~\ref{fig:1}(b). All parameters are given in cm$^{-1}$.}
\label{tab:1}
\centering
\begin{tabularx}{\linewidth}{XX}
\hline \hline
Parameter                            & Value (cm$^{-1}$) \\
\hline
$\Delta\epsilon$                    & 1500              \\
$V$                                & 300               \\
$\lambda$, $\omega_c$~(DL) & 540, 70           \\
$\lambda$, $\eta$, $\omega_0'$~(BO) & 161.6, 10, 1243 \\
\hline \hline
\end{tabularx}
\end{table}
\begin{lstlisting}[caption={\textbf{Dynamics of a Structured Bath Spin-Boson model}: Complete input file for running a spin-boson propagation with a structured bath.}, label={lst:full_file_example}]
# Import statements for simulation
from math import ceil
import os
import json as json
import numpy as np
from tqdm import tqdm
from tenso.prototypes.bath import gen_bcf
from tenso.prototypes.heom import system_multibath
# Import statement required to plot results
import matplotlib.pyplot as plt

# Generate the decomposed bath correlation function
# for HEOM propagation
bath_simulation = gen_bcf(
    re_d=[540], # Drude-Lorentz Reorganization energy (cm^{-1})
    width_d=[70], # Drude-Lorentz spectral density cut-off frequency (cm^{-1})
    freq_b=[1243], # Brownian spectral density frequency (cm^{-1})
    re_b=[161.6], # Brownian spectral reorganization energy (cm^{-1})
    width_b=[10], # Brownian spectral width (cm^{-1})
    temperature=300, # Temperature of this bath
    decomposition_method='Pade',
    n_ltc=1,) # Low temperature correction terms
end_time = 1000 # Propagation time in fs
dt = 1 # Time step size in fs
wfn = np.array([0.0, 1.0], dtype=np.complex128) # Initial wavefunction
out="tutorial_1" # Base name for the output

# Generate the propagator to evolve the system density matrix
propagator = system_multibath( 
    fname=out, # Specify output file
    init_rdo=np.outer(wfn, wfn.conj()), # Set system initial condition
    sys_ham=np.array([[-750.0,300.0],[300.0,750.0]], dtype=np.complex128), # Set the system Hamiltonian (cm^{-1})
    sys_ops=[np.array([[-0.5,0.0], [0.0,0.5]], dtype=np.complex128)], # Set the H_{SB} system-bath coupling operator(s) 
    bath_correlations=[bath_simulation], # Specify bath correlation function(s) generated by gen_bcf
    dim=25, # Hierarchy depth
    end_time=end_time, # End time of the simulation
    step_time=dt,) # Time step size
# Perform the dynamics simulation and generate a progress bar 
progress_bar = tqdm(propagator,total=ceil(end_time/dt))
for _t in progress_bar:
    progress_bar.set_description(f'@{_t: .2f} fs')
    
# Load results from the output file into a local array
results = np.loadtxt(out+".dat.log",dtype=np.complex128)
# Plot the populations of the two state of the spin-boson 
plt.plot(np.real(results[:,0]),np.real(results[:,1]),np.real(results[:,0]),np.real(results[:,4]))
plt.xlabel("Time (fs)")
plt.ylabel("Population")
plt.savefig(out+'.png')
plt.show()
\end{lstlisting}
On completion, this calculation will display a figure of the populations of the two states and save the image to a file. The output file, tutorial\_1.dat.log will record the complex values of all entries in the density matrix of the evolving system and output them in the order $\rho_{1,1} \; \rho_{1,2} ... \; \rho_{n,n-1}, \; \rho_{n,n}$ for each simulation time step. The dynamics are shown in Fig. \ref{fig:1}.

\subsubsection{Comparison of Different Tensor Decompositions}

\noindent
\texttt{TENSO} incorporates two built-in TTN decomposition schemes, the Tensor-Train (TT) decomposition and the balanced tensor tree (BTT) decomposition. The binary BTT is the default option and will be appropriate for most uses. The first example in Listing \ref{lst:full_file_example} addressed a spin-boson system using the BTT. We now demonstrate that a TT decomposition produces equivalent results.
To produce an equivalent calculation with a TT decomposition, we add the line \lstinline{frame_method= 'train',} in Listing \ref{lst:full_file_example} as an extra argument to \lstinline{system_multibath}. This is the only needed modification.
\begin{figure}[htb!]
    \centering
    \includegraphics[width=0.95\linewidth]{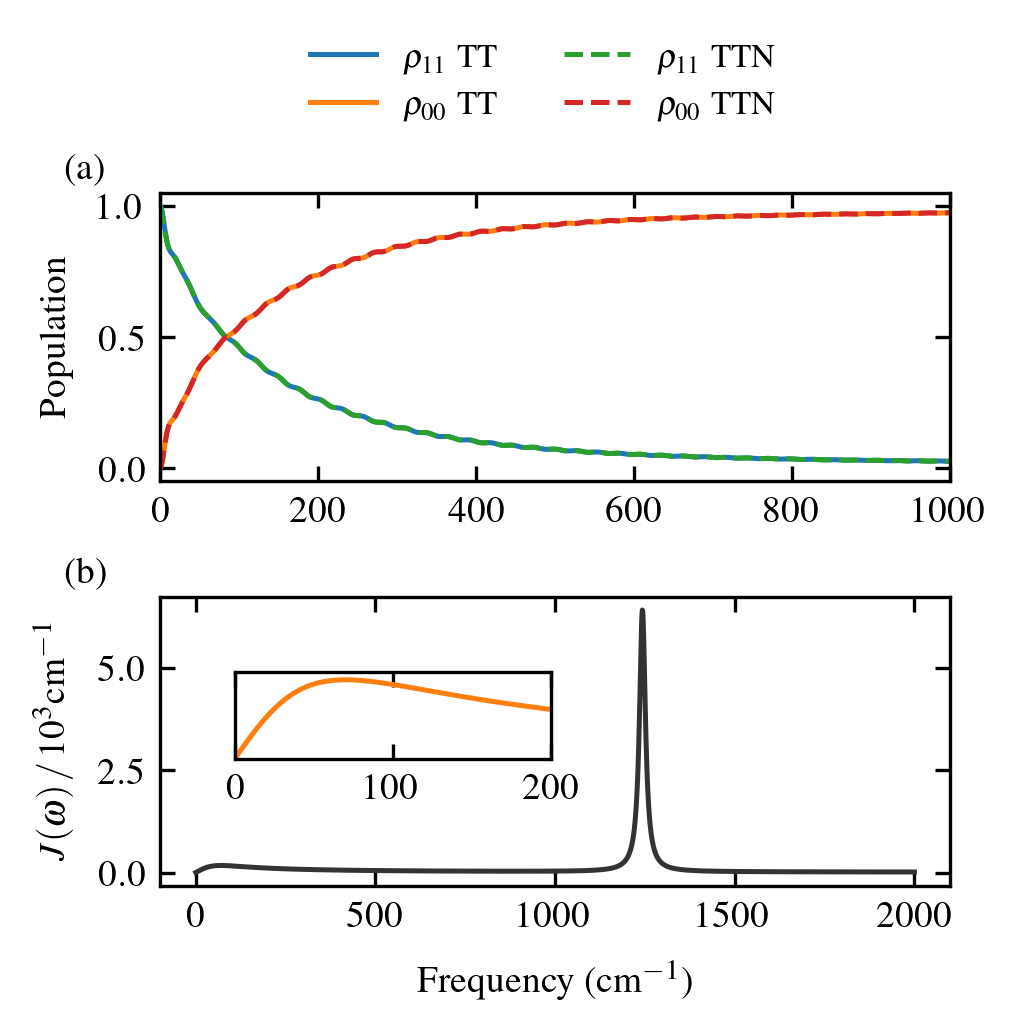}
    \caption{\textbf{Structured Spin-Boson Quantum Dynamics.} (a) Population relaxation dynamics of the spin-boson system, with the excited state in blue and ground state in orange, showing that dynamics are identical whether the TT or BTT tensor network decomposition is employed. (b) Structured spectral density employed in the simulation.}
    \label{fig:1}
\end{figure}

Fig.~\ref{fig:1} shows results of both the TT and BTT simulations of the spin-boson problem, demonstrating that they produce equivalent results. The dynamics show decay of the population and small oscillations attributed to interaction with the underdamped Brownian mode whose presence is very notable in the spectral density, Fig.~\ref{fig:1}(b). Note that all of \texttt{TENSO}'s convergence parameters controlling accuracy and expense, which will be discussed in depth in section \ref{sec:conpar}, are identical for these calculations. In this particular case, the convergence and performance of the simulation is not significantly impacted by the chosen tensor configuration.

\subsubsection{Non-Commuting Fluctuations}
\noindent
\texttt{TENSO}, as it is an HEOM-based method, enables modeling of open quantum systems coupled to non-commuting fluctuations. In this situation, an open quantum system is coupled to two or more distinct, independent thermal baths. Specifically, with the superscript $\textrm{b}$ referencing distinct baths, $H_{\mathrm{SB}} = \sum_\textrm{b} Q_\mathrm{S}^{(\textrm{b})} \otimes X_{\mathrm{B}}^{(\textrm{b})}$,  ${H}_\textrm{B} = \sum_{j,\textrm{b}} \left( \frac{(p^{(\textrm{b})}_j)^2}{2m^{(\textrm{b})}_j} + \frac{m^{(\textrm{b})}_j (\omega^{(\textrm{b})}_j)^2 (x^{(\textrm{b})}_j)^2}{2} \right)$, and the $Q_\mathrm{S}^{(\textrm{b})}$ do not commute. These systems may display interesting phenomena, including the apparent suppression of spin relaxation by decoherence\cite{garwola2024,palm2021} but their treatment can be challenging.\cite{garwola2024,palm2021,gribben2022,palm2018} As many systems relevant to quantum information science involve non-commuting fluctuations, including systems interacting simultaneously with radiation and phonon fields, addressing them is an important capability of \texttt{TENSO}.

\begin{figure}[htb!]
    \centering
    \includegraphics[width=0.95\linewidth]{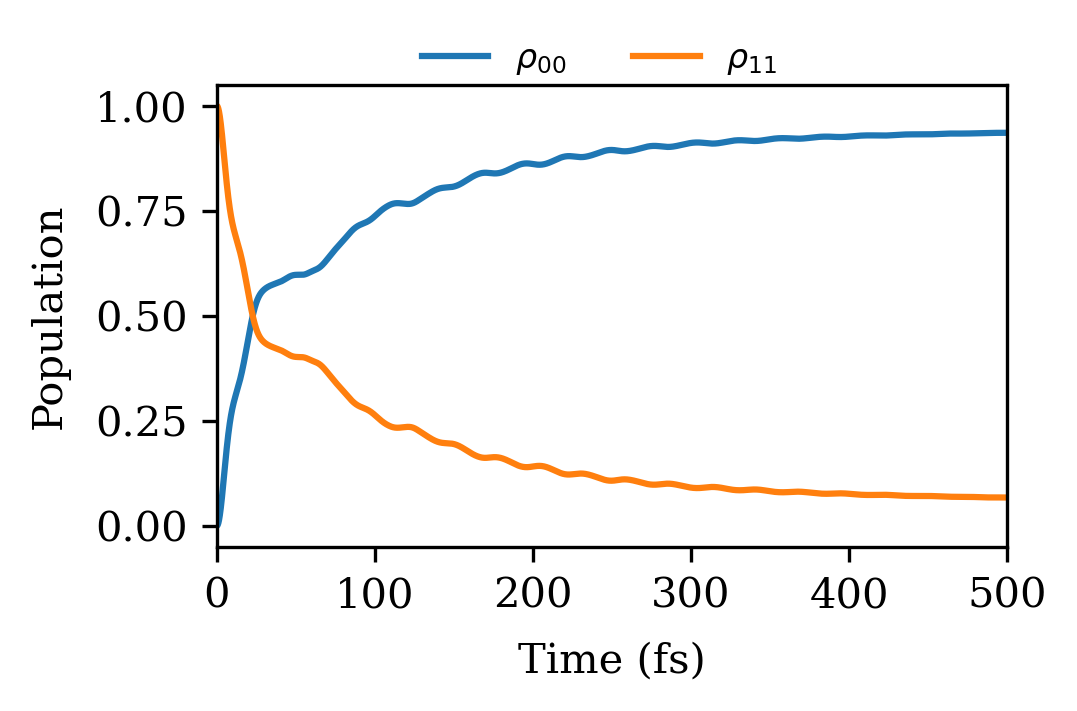}
    \caption{\textbf{Spin-Boson With Two Baths:} Evolution of a spin-boson system coupled to two non-commuting fluctuations.}
    \label{fig:nc_sb}
\end{figure}

A simple example of a system coupled to non-commuting fluctuations is a spin-boson problem where $H_{\mathrm{SB}} = \frac{1}{2}  \sigma_x \otimes X_{\mathrm{B}}^{(1)} + \frac{1}{2} \sigma_z \otimes X_\mathrm{B}^{(2)}$ . We calculate dynamics of the system specified in Listing \ref{lst:full_file_example} and in Table \ref{tab:1} with addition of a second, identical bath with $Q_\mathrm{S}^{(2)} =\frac{1}{2} \sigma_x$. This requires changing the propagator as shown in Listing \ref{lst:noncommute}.
\begin{lstlisting}[caption={\textbf{Non-Commuting Fluctuations}: Code modifications needed to set up a propagator when the system is coupled to two baths by non-commuting operators.}, label={lst:noncommute}]
# Set up the two system-bath coupling operators
bath_op1 = np.array([[-0.5,0.0],[0.0,0.5]],dtype=np.complex128)
bath_op2 = np.array([[0.0,0.5],[0.5,0.0]],dtype=np.complex128)

propagator = system_multibath(
    fname=out, 
    init_rdo=np.outer(wfn, wfn.conj()), 
    sys_ham=np.array([[-750.0,300.0],[300.0,750.0]], dtype=np.complex128), 
    sys_ops=[bath_op1,bath_op2], # Set the H_{SB} system-bath coupling operators 
    bath_correlations=[bath_simulation, bath_simulation], # Specify two identical correlation functions
    dim=30, 
    end_time=end_time, 
    step_time=dt,) 
\end{lstlisting}
The dynamics of the system coupled to the two non-commuting fluctuations, shown in Fig. \ref{fig:nc_sb}, differ significantly from those of the system coupled to a single bath in Fig. \ref{fig:1}. The total reorganization energy has doubled with the inclusion of the second bath, leading to faster relaxation. The apparent slowing of relaxation at approximately $50 \; \mathrm{fs}$ appears due to the competing influences of the two baths interacting with the system. 

\subsubsection{Driven Structured Spin-Boson Model}
\noindent
\texttt{TENSO} supports time-dependent Hamiltonians, enabling the simulation of driven phenomena in the presence of decoherence, such as photoexcitation of molecules in the condensed phase, spectroscopy, quantum control and quantum information processing.

Here, we simulate a spin-boson model resonantly driven by a $\frac{\pi}{2}$-pulse. In the absence of decoherence, this control pulse implements a perfect Hadamard gate on this qubit, transitioning it from the ground state $|0\rangle$ to $\frac{1}{\sqrt{2}} \left(|0\rangle + |1\rangle \right)$.  The total system Hamiltonian is given by
\begin{equation}
    H_{\text{S}} = \sigma_z \frac{\Delta\epsilon}{2} - \sigma_x \mu_0 E(t) 
\end{equation}
where $\mu_0$ is the transition dipole moment and $E(t)$ is the time-dependent laser field. 
The driving is realized by a single \(\pi/2\)-area Gaussian laser pulse, determined from the area theorem,\cite{mccall1967}
\begin{equation}
E(t) = E_0 \exp\left[-\frac{1}{2} \left(\frac{t - \tau}{\sigma}\right)^2 \right] \cos(\omega_L t),
\end{equation}
where \(E_0\) is the peak electric field amplitude, $\omega_L$ is the driving frequency, and \(\sigma\) is related to the pulse full width at half maximum (FWHM) by \(\sigma = \mathrm{FWHM} / \left( 2\sqrt{2\ln{2}}\right) \). In our simulations, FWHM corresponds to a value of 2.1233045~fs.
The system-bath coupling corresponds to pure dephasing noise, which is appropriate in many quantum computing investigations where dephasing is often  much faster than dissipation.

This simulation is carried out in \texttt{TENSO} by defining the laser parameters and a function \lstinline{laser_field_hadamard(t)} which returns the value of $-\mu_0E(t)$ with a maximum value of 3133.625~$\rm cm^{-1}$. 
The parameters for the bath are the same as those for the previous example outlined in Table \ref{tab:1} and the BCF portion of the input file requires no changes. The propagator must be modified to specify the time-dependent function and its associated operator, $\sigma_x$. The necessary changes to the previous example code are shown in Listing \ref{lst:driven}
\begin{lstlisting}[caption={\textbf{Time-Dependent Hamiltonians}: Code for time-dependent simulations. Here, the time-dependent drive is defined as a python function and the parameters \texttt{td\_f} and \texttt{td\_op} are passed to the \texttt{system\_multibath} propagator.}, label={lst:driven}]
# Final laser field parameters after unit conversions
t0_fs = 6.0
omega_rad_fs = 0.2825477
# The dipole has been incorporated into this value
interaction_max_cm = 3133.625
sigma_fs = 2.1233045
# Laser field function
def laser_field_hadamard(t):
    #Returns the laser field interaction energy at time t (fs)
    envelope = np.exp(-0.5 * ((t - t0_fs) / sigma_fs)**2)
    # Unitless argument inside cos: (rad/fs) * fs = radians
    carrier = np.cos(omega_rad_fs * t)
    return -interaction_max_cm * envelope * carrier
    
end_time = 40.0 # fs
dt = 0.05 # fs
wfn = np.array([0.0, 1.0], dtype=np.complex128) # Start in Ground State |0> (convention [exc, gnd])
out = 'laser_example'
propagator = system_multibath(
    fname='out',
    init_rdo=np.outer(wfn, wfn.conj()),
    # System Hamiltonian
    sys_ham=np.array([[750, 0.0], [0.0, -750]], 
                     dtype=np.complex128),
    # System-Bath coupling operator (Sigma_z / 2)
    sys_ops=[np.array([[0.5, 0.0], [0.0, -0.5]], 
                    dtype=np.complex128)],
    bath_correlations=[bath_simulation],
    # Time-dependent driving function & Operator
    td_f=laser_field_hadamard,
    td_op=np.array([[0.0, 1.0], [1.0, 0.0]],
                   dtype=np.complex128), #(sigma_x)
    dim=30,
    end_time=end_time,
    step_time=dt,
)

\end{lstlisting}
\begin{figure}[htb!]
    \centering
    \includegraphics[width=0.95\linewidth]{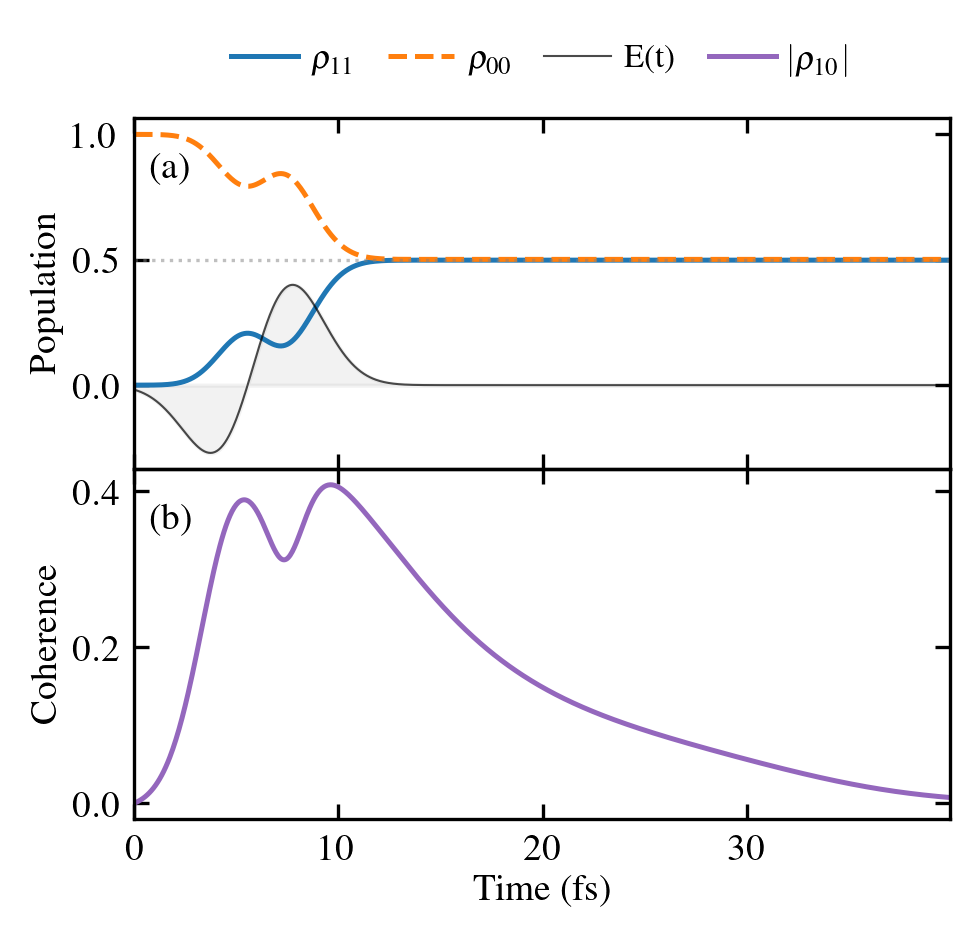}
    \caption{\textbf{Driven Spin-Boson Dynamics:} (a) Time evolution of populations and (b) coherence evolution under a resonant $\pi/2$ Gaussian pulse while interacting with a structured bath. (c) The structured spectral density of the thermal bath. The laser-induced control dynamics demonstrates coherent excitation in the two-level system.}
    \label{fig:driven_spinboson}
\end{figure}

In Fig. \ref{fig:driven_spinboson} the system's state is driven to a superposition, however this half rotation in the Bloch sphere is not complete. The system state never reaches $\frac{1}{\sqrt 2}(|0\rangle + |1\rangle)$ and the coherences decay quickly due to dephasing, meaning that a noisy Hadamard gate has been modeled. This demonstrates \texttt{TENSO}'s capability to simulate the impact of structured environmental noise on driven quantum systems, a situation relevant to many quantum control and quantum computing problems.

\subsubsection{Adaptability of the Framework to Other Equations of Motion}
Although the principal application of \texttt{TENSO} and the principal focus of this tutorial is HEOM, the tensor network framework underlying the program is easily adapted to other equations of motion with propagators in sum-of-products form. In particular, this is demonstrated by the implementation of thermofield strategy using ML-MCTDH, which is a tensor-based method for solving the time-dependent Schr\"odinger equation, in TENSO.\cite{chen2026} The thermofield strategy allows the wavefunction-based MCTDH to simulate finite temperature dynamics by mapping them to zero temperature dynamics in an extended Hilbert space.\cite{takahashi1996,tamascelli2019,takahashi2025} Due to the multi-layered tree structure, the MCTDH implementation constitutes the ML-MCTDH method.

Performing a MCTDH calculation requires discretizing the bath, and the situation in Fig.~\ref{fig:1} requires a large discretization and significant computational resources. The technical aspects of bath discretization, as required to use MCTDH for thermal dynamics using TEDOPA and related methods, are detailed in Refs. \onlinecite{chen2026}, \onlinecite{tamascelli2019}, \onlinecite{de2024} and \onlinecite{vega2015}. For the purposes of this demonstration, a system with a single Brownian spectral density which does not require as many computational resources to achieve MCTDH convergence has been selected to compare to HEOM. The required changes to the simulation code to perform MCTDH are demonstrated in Listing \ref{lst:mctdh}. The HEOM modules are replaced with MCTDH modules and \lstinline{gen_bcf} is replaced by \lstinline{gen_star_boson} to perform the MCTDH discretized bath decomposition. The overall changes required to perform the simulation are minimal, demonstrating the versatility of \texttt{TENSO}'s implementation framework and usage. The HEOM comparison for this example is generated by adjusting Listing \ref{lst:full_file_example} so that the bath contains only the Brownian mode specified in Listing \ref{lst:mctdh}. MCTDH calculations and their efficiency in comparison to HEOM calculations in \texttt{TENSO}'s tensor network approach are largely beyond the scope of this tutorial and interested readers may find an in depth discussion in Ref.~\onlinecite{chen2026}. 

\begin{figure}[htb!]
    \centering
    \includegraphics[width=0.95\linewidth]{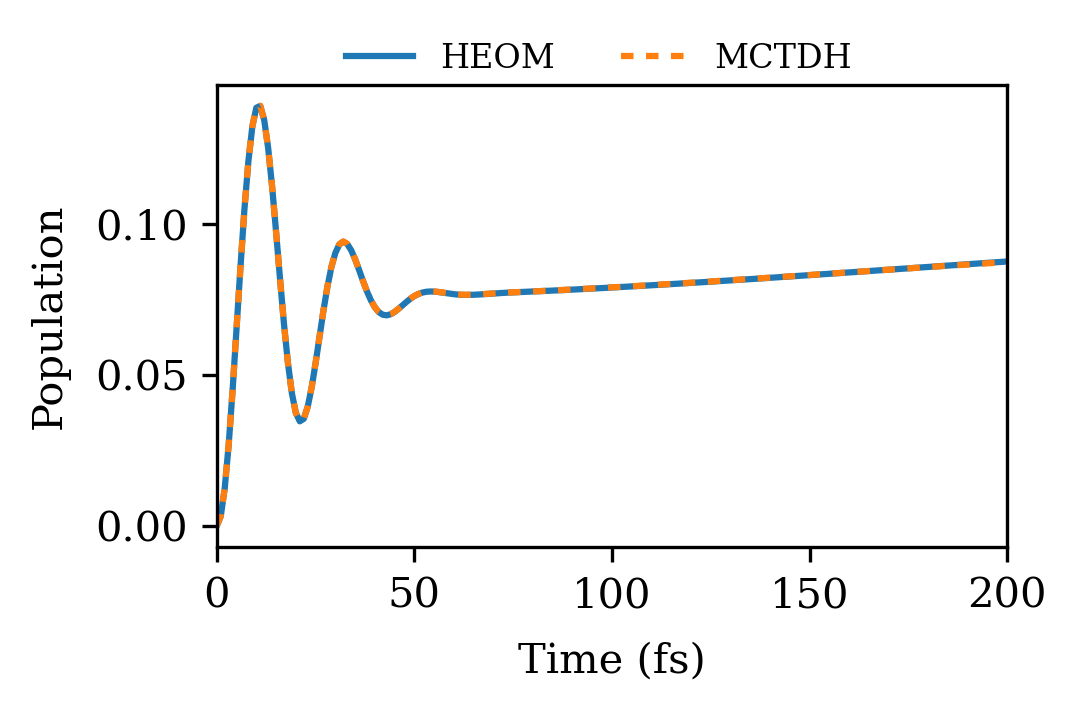}
    \caption{\textbf{MCTDH vs HEOM.} Short comparison of an MCTDH and HEOM calculation for a simple Brownian bath at 300 ~K. The HEOM calculation was carried out with \lstinline{n_ltc=2}, \lstinline{dim=20}, and \lstinline{dt=1} with the default mixed propagation strategy.} 
    \label{fig:mctdh_figure}
\end{figure}

Despite our focus on HEOM rather than MCTDH, we would be remiss not to offer a general word on convergence. The discretization of the bath in MCTDH inevitably leads to recurrence and inaccuracy after a finite propagation time, which can be mitigated by increasing \lstinline{n_discretization} but can never be escaped entirely. In Fig.~\ref{fig:mctdh_figure}, the selected discretization is sufficient for 200 fs, but not for 300 fs.
\begin{lstlisting}[caption={\textbf{MCTDH Versus HEOM}: Code for a MCTDH calculation with a Brownian spectral density to compare to HEOM.}, label={lst:mctdh}]
from math import ceil
import numpy as np
from tqdm import tqdm
# Import MCTDH module rather than HEOM modules
from tenso.prototypes.mctdh import system_multibath
from tenso.prototypes.bath import gen_star_boson

out = "mctdh"
# Generate the type of bath discretization required for MCTDH
bath_simulation = gen_star_boson(
    re_b=[200],
    width_b=[10],
    freq_b=[50],
    temperature=300,
    cutoff=200, # Maximum frequency in the bath discretization
    n_discretization=50, # Number of modes in the discretization
    discretization_method='TEDOPA', # Method of choosing discrete modes
)

end_time = 200
dt = 1
wfn = np.array([0.0, 1.0], dtype=np.complex128) # Initial wavefunction

propagator = system_multibath(
    fname=out,
    init_wfn=wfn, # The propagator requires an initial wavefunction, not density matrix
    sys_ham=np.array([[-750.0,300.0],[300.0,750.0]], dtype=np.complex128),
    sys_ops=[np.array([[-0.5,0.0], [0.0,0.5]], dtype=np.complex128)],
    baths=[bath_simulation], # Note the different name for the bath input
    end_time=end_time,
    dim=8, # Dimension of each mode in the discretization
    step_time=dt,)
progress_bar = tqdm(propagator,total=ceil(end_time/dt))
for _t in progress_bar:
    progress_bar.set_description(f'@{_t: .2f} fs')


\end{lstlisting}

\subsubsection{Convergence Parameters}
\label{sec:conpar}
\texttt{TENSO} has several parameters that control the numerical accuracy of the propagation including the time step size, propagation method, rank of the tensor network, hierarchy depth, and number of low temperature corrections included in the BCF decomposition. An increase in computational accuracy inevitably brings higher computational cost.

The simplest adjustment parameter is time step size. An initial step size which is too large may result in numerical problems or simulation failure. The initial step size is set by the parameter \lstinline{step_time} in the propagator. A good rule of thumb is that it should be at most $1/20$th of the fastest timescale in the problem.

The default propagation method in \texttt{TENSO} begins with the use of  the ps2 projection splitting method that adapts the TTN ranks to the problem at hand and therefore has a variable memory footprint. \texttt{TENSO} then transitions to a static rank, constant memory footprint direct integration method, vmf. This scheme should be appropriate for most uses. More information on adjusting the propagation method and error tolerances is located in Sections \ref{sec:defaults} and \ref{sec:prop_params}.

\begin{figure}[htb!]
    \centering
    \includegraphics[width=0.95\linewidth]{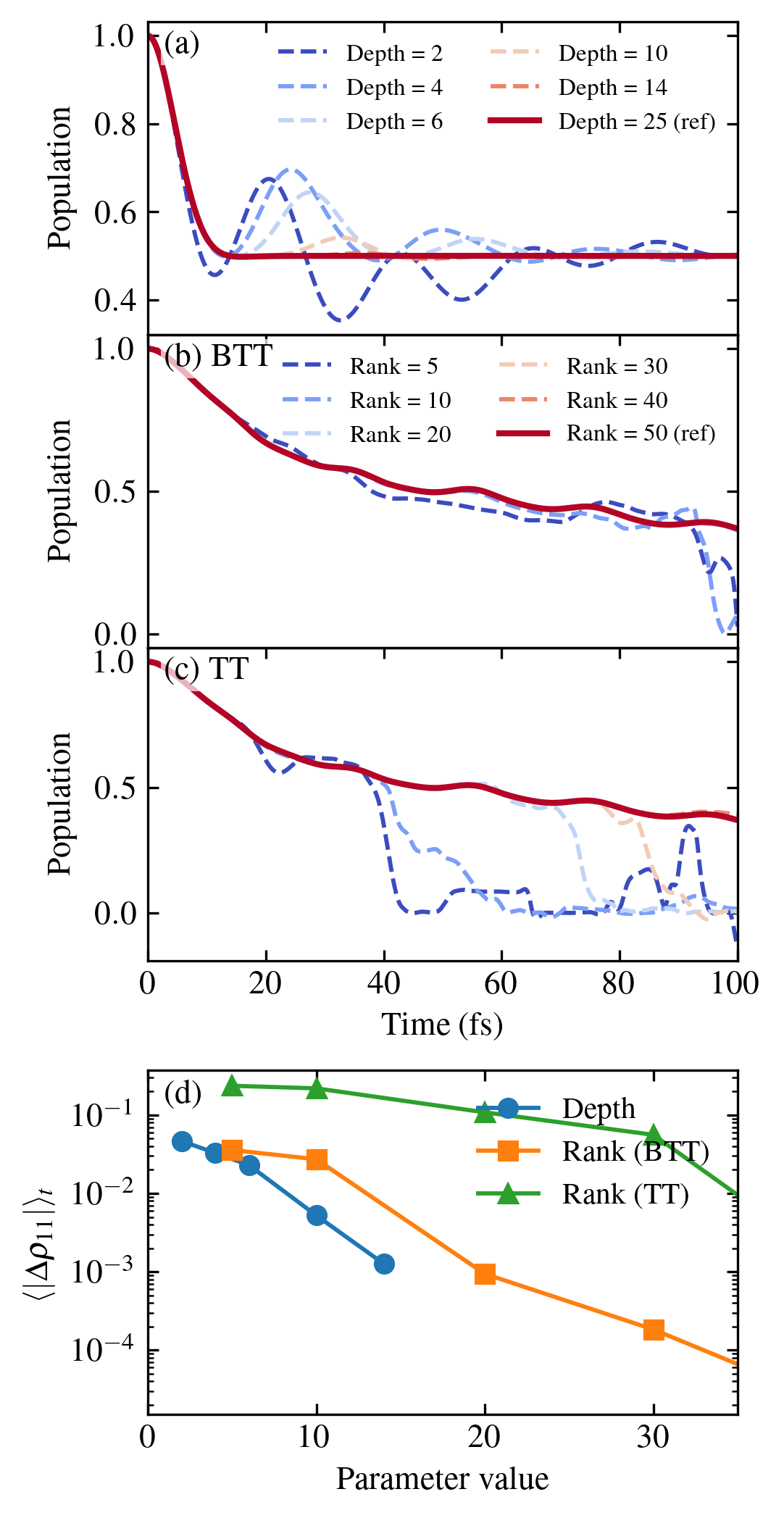}
    \caption{\textbf{Convergence Demonstration.} (a) Population of a spin-boson model simulated with the BTT method for several different hierarchy depths demonstrating the spurious oscillations that occur when the depth is insufficient. (b) Population of a spin-boson model simulated with the BTT method for several different ranks showing less sensitivity to rank. (c) Population of a spin-boson model simulated with the TT method for several different ranks showing more sensitivity to rank. (d) Comparison of mean absolute deviation of population from reference population with depth and rank.}
    \label{fig:converging}
\end{figure}

We will demonstrate convergence of calculations with respect to depth and rank using spin-boson models specified in Table \ref{tab:2}.  The first model in Fig. \ref{fig:converging} (a) employs a symmetric model ($\Delta\epsilon = 0$) with a purely DL bath, while (b) and (c) use an asymmetric system coupled to a combined DL and BO bath. The system Hamiltonian and system-bath coupling operator are the same as in Listing~\ref{lst:full_file_example}.

Truncation depth of the hierarchy expansion for each feature is controlled by the parameter \lstinline{dim} in the propagator specifications. The default value is $5$, but different models will require different depths. HEOM, when performed with an inadequate hierarchy depth, may produce dynamics with spurious oscillations or revivals\cite{cainelli2021,Krug2023} so it is crucial to check convergence with respect to hierarchy depth.\cite{Ishizaki2005} This issue is demonstrated in Fig.~\ref{fig:converging} (a), which performs the same calculations as in Listing \ref{lst:full_file_example} with a few modifications to the parameters. Only the DL term is used in the spectral density, $\Delta \epsilon=0$, and the system-bath coupling operator is ${H_\textrm{S}}= \sigma_x \otimes {X_B}$. The truncation depth is set to \lstinline{dim=n} for each of $n=2,4,6,10,14,25$. Oscillations in the population when $n$ is less than $25$, whether very dramatic at $n=2$ or minimal at $n=14$, are the result of insufficient hierarchy depth, but superficially resemble coherent dynamics that may occur in other quantum systems.
Similar spurious oscillations can occur when insufficient low temperature corrections are included,\cite{Ishizaki2005} but this situation may produce not only spurious oscillations but physically unreasonable dynamics such as negative populations.\cite{dunn2019,fay2022} The number of low temperature correction terms included during the bath decomposition is determined by \lstinline{n_ltc} during BCF construction. The default parameter is $0$. The number required depends on the specific system and bath energy scales. For example, in Listing \ref{lst:full_file_example}, \lstinline{n_ltc=1} is sufficient. Note that the HEOM requires increasingly more features in the environment as the temperature is lowered and can become numerically intractable.

The tensor network rank is also a critical convergence parameter governing the accuracy of the simulations. Figures~\ref{fig:converging} (b) and (c) display the convergence of the population dynamics with respect to the bond rank for the BTT and TT decompositions, respectively, in a case where $H_{\mathrm{SB}} = \sigma_z \otimes X_\mathrm{B}$. The simulations use the fixed rank ps1 as the propagation strategy. 
The BTT structure in (b) exhibits quick convergence towards the reference solution (Rank = 40), whereas the TT decomposition in (c) requires significantly higher ranks to achieve comparable accuracy and displays unphysical artifacts at low ranks.
The time-averaged absolute error versus the reference simulation, $\langle|\Delta\rho_{11}|\rangle_t = \frac{1}{N_\mathrm{ts}}
\sum_{k=1}^{N_{\mathrm{ts}}}|\rho_{11}(t_k) - \rho_{11}^{\mathrm{ref}}(t_k)|$,  where $N_{\mathrm{ts}}$ is the number of time steps in the simulation, is shown in Fig.~\ref{fig:converging} (d) as a function of the parameter value for all three cases, providing a quantitative measure of convergence. The BTT rank converges most rapidly, reaching errors below $10^{-3}$ at moderate rank values, while the TT decomposition and hierarchy depth require larger parameter values to achieve similar accuracy. 

Additional discussion of convergence testing is found in Section \ref{sec:defaults}.

\begin{table}[htb]
\caption{Simulation parameters for the spin--boson models used in the
convergence tests of Fig.~\ref{fig:converging}. All parameters are given
in cm$^{-1}$.}
\label{tab:2}
\centering
\begin{tabularx}{\linewidth}{XXX}
\hline \hline
Parameter & Panel (a) (cm$^{-1}$) & Panel (b,c) (cm$^{-1}$) \\
\hline
$\Delta\epsilon$ & 0 & 1500 \\
$V$ & 300 & 300 \\
$\lambda$, $\omega_c$ (DL) & 540, 70 & 540, 70 \\
$\lambda$, $\eta$, $\omega_0'$ (BO) & 0, 0, 0 & 330, 4, 1663 \\
\hline \hline
\end{tabularx}
\end{table}

\subsection{Example 2: FMO Complex With a Structured Environment}
\noindent
The Fenna--Matthews--Olson\cite{fenna1975,fenna1979} complex is employed in studies of excitonic energy transfer, coherence and entanglement in photosynthetic complexes, and for benchmarking new methods.\cite{milder2010,gonzalez-soria2019,kell2019,bose2023,jang2018}  Studies have examined non-Markovian effects and the impact of different thermal baths on the FMO complex to improve our understanding of photosynthetic processes.\cite{rodriguez2025,chenu2015,hu2022} We will use the FMO complex to demonstrate \texttt{TENSO}'s utility in addressing larger systems with multiple baths and complicated descriptions of the environment spectral density involving three or more Brownian terms. 

In the site basis, the FMO system Hamiltonian is
\begin{equation}
    {H}_\mathrm{S} = \sum_i \epsilon_i |i\rangle \langle i| + \sum_{i \neq j} V_{ij} \left( |i\rangle \langle j| + |j\rangle \langle i| \right),
\end{equation}
where $\epsilon_i$ denote site energies, and $V_{ij}$ the couplings between sites $i$ and $j$. The FMO complex is usually described by seven or eight sites, but we will use a reduced three site description to control computational cost. The three site model captures major features effectively\cite{schulze2015} and will suffice for the purpose of our demonstration. 

The matrix representation of the three site Hamiltonian in $\text{cm}^{-1}$ is\cite{Ishizaki2009a}
\[{H}_\mathrm{S}=
\begin{bmatrix}
    200 & -87.7 & 5.5\\
    -87.7 & 320 & 30.8\\
    5.5 & 30.8 & 0
\end{bmatrix}.
\]
Most HEOM simulations of the FMO complex employ an independent bath with a Drude--Lorentz spectral density at each site. While this captures significant dynamical features, higher frequency modes—which substantially influence short-time behavior—are typically underestimated;\cite{lorenzoni2025} thus simulations with more accurate spectral densities are desirable.\cite{kim2018, maity2023} \texttt{TENSO} makes studies incorporating structured spectral densities for this system computationally feasible for HEOM. To reduce the number of low temperature corrections, high frequency modes such as those encountered in the FMO model can be addressed by delta function treatment.\cite{Ishizaki2005} This method is compatible with \texttt{TENSO} but not currently implemented.

The system-bath coupling Hamiltonian is
\begin{equation}
    {H}_\mathrm{SB} = \sum_\textrm{b} \sum_i \hbar \omega_{\textrm{b},i} \sqrt{2 S_{\textrm{b},i}}|i\rangle \langle i|\otimes X_B^{(\textrm{b})},
\end{equation}
with $S_{\textrm{b},i}$ a Huang-Rhys factor and $\omega_{\textrm{b},i}$ the $i$th oscillator frequency of bath $\mathrm{b}$. Note that each site is coupled to an independent bath. Our model approximates the spectral density via a sum of six underdamped Brownian oscillator terms which approximate one recently determined from spectroscopic measurements.\cite{lorenzoni2024}

\begin{figure}[htb!]
    \centering
    \includegraphics[width=0.95\linewidth]{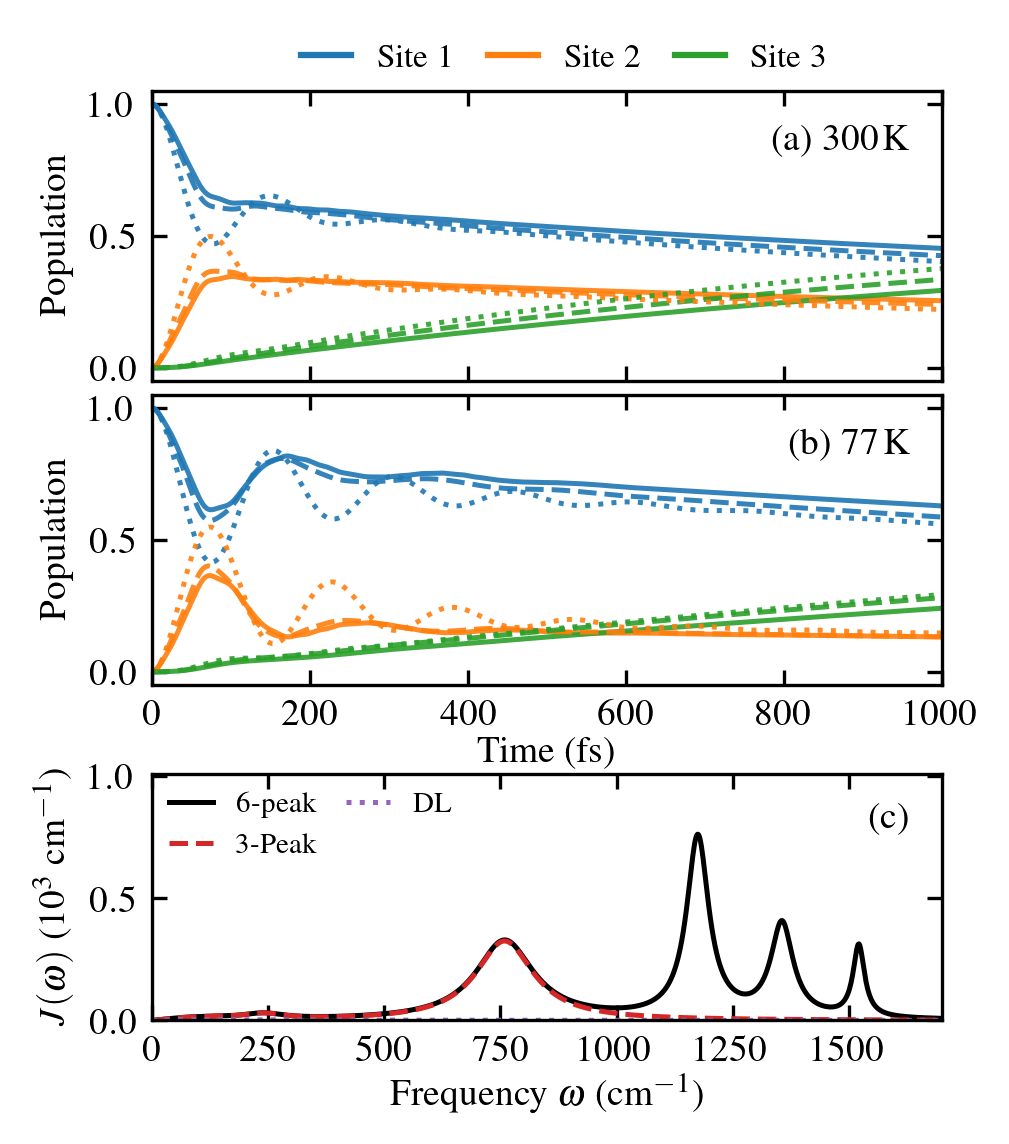}
    \caption{\textbf{Population Dynamics in the FMO Complex.} Population evolution at (a) $T = 300$~K and (b) $T = 77$~K for three sites, computed with the BTT tensor network decomposition for three different structured spectral densities (solid, dashed, and dotted lines). (c) The three structured spectral densities employed in the simulations.}
    \label{fig:FMO_populations}
\end{figure}

Listing \ref{lst:fmo} contains the entire input file necessary to run the FMO example, excluding the import statements, which are identical to Listing \ref{lst:full_file_example}. Note that a different propagation method is used in this simulation. Propagators are discussed more in Sections \ref{sec:defaults} and \ref{sec:prop_params}. We will make additional small modifications to this input file to compare the FMO dynamics with different thermal baths. First, we will run calculations at both 300~K and 77 ~K and use three low temperature corrections, \lstinline{n_ltc=3}, for all 77~K calculations. Second, we will employ two simpler spectral densities, one using only the first three Brownian peaks, the three located below $1000 \; \mathrm{cm}^{-1}$, and one using a simple Drude-Lorentz spectral density with $\lambda^{\mathrm{DL}}=35 \; \mathrm{cm}^{-1}$ and $\omega_c^{\mathrm{DL}}=106.18 \;\mathrm{cm}^{-1}$.\cite{Ishizaki2009a} These three spectral densities are shown in Fig.~\ref{fig:FMO_populations} (c).
\begin{lstlisting}[caption={\textbf{Three-site FMO quantum dynamics:} Input file for the simulation of the FMO complex. The ordering inside the \texttt{bath\_correlations} list must match the ordering in the \texttt{sys\_ops} list so that there is a one-to-one correspondence between system operators and baths.}, label={lst:fmo}]
bath = gen_bcf(
    re_b=[26.24, 13.832, 101.479, 57.575, 25.764, 9.126],
    freq_b=[160, 247, 763, 1175, 1356, 1521],
    width_b=[133, 53, 76, 29, 29, 15],
    temperature=300,
    decomposition_method='Pade',
    n_ltc=1, # 3 needed at 77 K
)

h = np.array([
    [200, -87.7, 5.5],
    [-87.7, 320, 30.8],
    [5.5, 30.8, 0]
], dtype=np.complex128)

sys_ops = []
end_time = 1000.0
dt = 0.05
# Add a coupling operator for each site
for i in range(3):
    op_i = np.zeros((3, 3))
    op_i[i, i] = 1.0
    sys_ops.append(op_i)

wfn = np.zeros(3)
wfn[0] = 1.0  # Initial state localized at site 1
out = "3_level_FMO_structured"
propagator = system_multibath(
    fname=out,
    init_rdo=np.outer(wfn, wfn.conj()),
    sys_ham=h,
    sys_ops=sys_ops,
    # Couple an identical, independent bath to each site
    bath_correlations=[bath, bath, bath],
    end_time=end_time,
    step_time=dt,
    # Use a static rank projector splitting propagation method
    ps_method="ps1",
    dim=20,
)

\end{lstlisting}
In Fig.~\ref{fig:FMO_populations}, we show the dynamics of the FMO complex under the six different baths. Population evolution with the simple Drude-Lorentz bath corresponds to dotted lines, solid lines correspond to the structured spectral density with all six Brownian peaks and the dashed lines denote the spectral density which uses only the three lowest frequency Brownian peaks. Significantly different population dynamics are observed for the three different spectral densities, with these differences being far more pronounced at lower temperature. Higher temperature suppresses population oscillations associated with coherent effects, an expected outcome.\cite{duan2017} The Drude-Lorentz spectral density produces very divergent dynamics in comparison to the more structured cases which, notably, have a higher total reorganization energy. For the structured spectral density selected, apparent coherent oscillations are much less pronounced than for the Drude-Lorentz spectral density.

The employment of more realistic, structured spectral densities can have a large impact on the dynamics of the FMO complex. \texttt{TENSO}'s ability to efficiently simulate systems with these features is a significant asset for treating photosynthetic complexes and related biological systems. The six Brownian peak spectral density at $77$~K yields  $K=15$ features (12 from the BO features and 3 from the low temperature corrections) per site or 45 total features,
which is significantly beyond the applicability of standard HEOM. Note that a thorough assessment of the appropriateness of this and other suggested structured spectral densities for the FMO complex is beyond the scope of this tutorial and this demonstration seeks only to emphasize their importance and \texttt{TENSO}'s ability to address them.

\subsection{Example 3: Entanglement Sudden Death}
\label{sec:esd}
Entanglement between quantum subsystems plays a foundational role in quantum information science and other fields of modern physics.\cite{horodecki2009,orszag2010,aolita2015} Long-lived, controllable entanglement between subsystems is a key resource in quantum communication, quantum sensing, and quantum computing.\cite{aolita2015,ursin2007,jozsa2003,guo2020} Entanglement has also become an important concept in quantum chemistry, offering new insights on electron correlations, molecular dynamics and open quantum systems. \cite{huang2006,hartmann2006,boguslawski2015}

However, as a quantum system interacts with its environment, decoherence occurs and entanglement between subsystems decays, presenting a major challenge for hardware design in quantum technologies.\cite{kjaergaard2019,chirolli2007,zhang2019} Density matrix coherence terms often decay exponentially depending on the interaction of an open quantum system with its environment but will remain finite at all times. Entanglement between subsystems, however, may decay to zero within finite time, a subject of intense research known as entanglement sudden death (ESD).\cite{yu2004,yu2009,mazzola2009,almeida2007,roszak2010,parez2024,nair2024}

The simplest system which can display meaningful entanglement is a pair of qubits, equivalently two spin-boson models. As the impact of bath models and non-Markovian effects are of keen interest in this simple model,\cite{wang2018,ficek2006,ma2012,jiang2018,bellomo2008} we use it as an example of the power of \texttt{TENSO} to address problems in quantum information sciences.

We consider two qubits in an initially maximally entangled state, $\left|\psi_0\right\rangle = \frac{1}{\sqrt{2}}(\left|01\right\rangle - \left|10\right\rangle)$, each interacting with a local Bosonic bath. The system Hamiltonian is given by
\begin{equation}
H_\mathrm{S}=\frac{\hbar \omega_1}{2}{\sigma_z}^{(1)}+\frac{\hbar \omega_2}{2}{\sigma_z}^{(2)}
\end{equation}
where $\sigma_z$ is the usual Pauli operator with the superscript indicating the qubit on which it operates and $\omega_i$ are the frequencies of the qubits. The interaction between the system and baths is given by
\begin{equation}
H_\mathrm{SB}=\hbar \sigma_x^{(1)}\otimes  X_\mathrm{B}^{(1)}+\hbar \sigma_x^{(2)}\otimes  X_\mathrm{B}^{(2)},
\end{equation}
where the collective coordinates of each bath are distinguished by superscripts.

We measure entanglement using concurrence. Concurrence is defined for a two-qubit density matrix $\rho$ using the spin-flipped state
\begin{equation}
\tilde{\rho}=(\sigma_y\otimes\sigma_y)\rho^*(\sigma_y\otimes\sigma_y) 
\end{equation}
where $\rho^*$ is the complex conjugate of $\rho$. We compute the eigenvalues $\lambda_1,\ \lambda_2,\ \lambda_3,\ \lambda_4$ of the matrix\cite{wootters2001}
\begin{equation}
R=\sqrt{\sqrt{\rho}\tilde{\rho}\sqrt{\rho}}
\end{equation}
and arrange them in decreasing order, $\lambda_1 \ge \lambda_2\ge \lambda_3\ge \lambda_4$ then find the concurrence $C(\rho)$ as
\begin{equation}
C(\rho)=\text{max}(0,\lambda_1-\lambda_2- \lambda_3- \lambda_4).
\end{equation}
If $C(\rho)>0$ the state is entangled and if $C(\rho)=0$ the state is separable.

To simulate the pair of qubits, the Hamiltonian and bath parameters for the spin-boson problem demonstrated in Listing \ref{lst:full_file_example} are modified as shown in Listing \ref{lst:concur}. The initial state will be maximally entangled. We will make additional modifications to bath specifications of the Drude-Lorentz bath which, unless otherwise noted, has $\lambda=50 \; \mathrm{cm}^{-1}$, $\omega_c=30\;\mathrm{cm}^{-1}$, and ${T} = 300$~K.
\begin{lstlisting}[caption={\textbf{Entanglement Sudden Death:} Python code for concurrence time evolution of a two qubit system.},label={lst:concur}]
# Define Pauli matrices
sigma_x = np.array([[0, 1], [1, 0]], dtype=np.complex128)
sigma_z = np.array([[1, 0], [0, -1]], dtype=np.complex128)
I = np.eye(2, dtype=np.complex128)
# Define parameters
omega = 20  
# Define single qubit Hamiltonian
H1 = (omega / 2) * sigma_z + omega/2 * I
# Construct two-qubit Hamiltonian without interaction
h = np.kron(H1, I) + np.kron(I, H1)

Q_s1=np.kron(sigma_x,I)
Q_s2=np.kron(I,sigma_x)

sys_ops = [Q_s1,Q_s2] #The system operators are sigma_x for both qubits 

rdo_0 = np.array([[0, 0, 0, 0],
                  [0, 0.5, -0.5, 0],
                  [0, -0.5, 0.5,0],
                  [0, 0, 0,0]],
                 dtype=np.complex128) #This is the initial state of the two-qubit system

propagator = system_multibath(
    fname=out,
    init_rdo=rdo_0,
    sys_ham=h,
    sys_ops=sys_ops,
    bath_correlations=[bath]*2,
    end_time=end_time,
    step_time=dt,
    dim=45
)

\end{lstlisting}
\begin{figure}[ht!]
    \centering
    \includegraphics[width=0.95\linewidth]{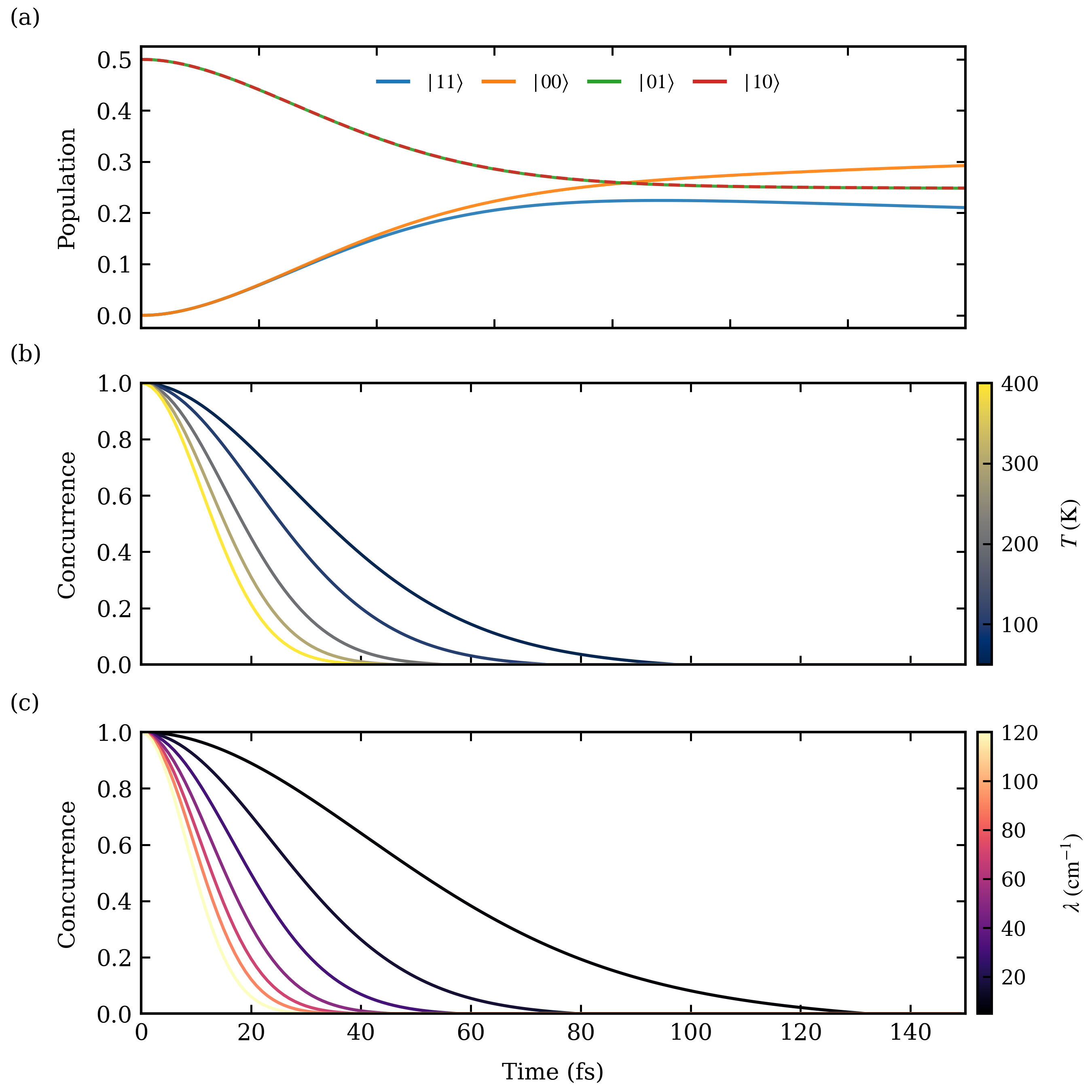}
    \caption{\textbf{Two-Qubit Quantum Dynamics and Entanglement Decay.} (a) Population dynamics of the two qubits at $T = 50$~K. (b) Decay of the concurrence between the two qubits for varying temperature, with the color scale ranging from low (blue) to high (yellow) values. (c) Decay of the concurrence under variation of the reorganization energy $\lambda$, with the color scale ranging from weak (blue) to strong (magenta) system--bath coupling.}
    \label{fig:qubits_pop}
\end{figure}
In Fig. \ref{fig:qubits_pop}, example population dynamics of the two qubits are displayed at 50 K where equilibration requires more than $150\;\mathrm{fs}$ to complete and populations are significantly biased in favor of the lower energy $|00\rangle$ state. In Fig.~\ref{fig:qubits_pop} (b) and (c), we see that entanglement survival time increases with a lower temperature bath and smaller reorganization energy. Physically, larger $\lambda$ indicates stronger system-bath coupling and environmental fluctuations, while a higher temperature $T$ amplifies thermal noise, both of which strengthen decoherence channels that rapidly suppress entanglement.

For the sake of expediency, the bath was kept simple in this tutorial. The impact of varying even simple bath parameters is generally notable, and \texttt{TENSO} allows for the systematic study of entanglement dynamics under more complicated, structured baths for more specialized applications in quantum information sciences.

\section{Settings Recommendations for Usual Problems}
\label{sec:defaults}
The convergence parameters and propagation methods required depend on the system under study. Here, we provide some general guidance for initial selections. \texttt{TENSO} can either run a calculation with one propagation strategy, requested by passing \lstinline{stepwise_method='simple'} to \lstinline{system_multibath}, or run an auxiliary method, specified by \lstinline{auxiliary_ps_method} for an initial period followed by a main method specified by \lstinline{ps_method}. We recommend initially addressing a new problem with the default mixed propagation strategy. This runs the adaptive rank projector splitting ps2 propagation as the auxiliary method to dynamically adjust ranks followed by fixed rank direct vmf propagation with the default adaptive time step fourth order Runge-Kutta method for the bulk of the calculation. This is the behavior of \texttt{TENSO} if no information about propagation method is specified.

In some cases, the projector splitting ps1 propagation method may be preferable to vmf as the main method. In the case that an adaptive rank is not desired, the default mixed propagation strategy with auxiliary method ps1 and main method vmf is recommended. This helps to avoid regularization errors which often occur with vmf at early times. Possible auxiliary and main method combinations are reviewed in Table \ref{tab:strats}.
\begin{table}[htb]
\caption{Comparison of available propagation strategies in \texttt{TENSO}. Mixed propagation starts with an auxiliary method then transitions to a main method. Simple propagation has no auxiliary method. Combinations not shown in this table are never recommended.}
\label{tab:strats}
\centering
\begin{tabularx}{\linewidth}{ccXX}
\hline \hline
Auxiliary & Main & Assessment & Recommendation  \\
\hline
ps2 & vmf & adaptive rank, adaptive time step allowed, avoids many vmf regularization errors & Recommended method (default) \\
\hline
ps2 & ps1 & adaptive rank, time step fixed & Recommended if adaptive timestep isn't required or vmf encounters stability problems\\
\hline
ps1 & vmf & fixed rank, adaptive timestep allowed, avoids many vmf regularization errors & Recommended if a fixed rank and adaptive time step is desired\\
\hline
None & ps2 & adaptive rank throughout calculation, usually high numerical cost & Not recommended\\
\hline
None & ps1 & fixed rank, fixed time step & Recommended if fixed rank is desired and adaptive time step is not needed\\
\hline
None & vmf & fixed rank, adaptive timestep, potential regularization errors & Not recommended\\
\hline \hline
\end{tabularx}
\end{table}
Table \ref{tab:defaults} reviews general recommendations for initial convergence parameter selections including guidance on how to tighten the parameter to confirm convergence of the calculation. Note that the fixed time step ps1 and ps2 methods require more careful convergence checking with respect to the size of the time step. Parameters not previously discussed include \lstinline{ode_rtol}, \lstinline{ode_atol}, \lstinline{ps2_atol}, \lstinline{ps2_ratio} and \lstinline{max_auxiliary_rank}. The first two control the relative and absolute tolerance of the ordinary differential equation integration. The last three control how many singular values are trimmed during the ps2 SVD steps and the maximum rank that ps2 will allow in the tensor network. The default parameters for these five should be well-suited to most applications, but can be adjusted as needed.

Ideally, convergence can be confirmed with a single calculation in which all of the parameters are relaxed, meaning hierarchy depth, low-temperature corrections and rank are slightly decreased while time step size is increased. If no significant changes are observed in the dynamics, convergence is likely achieved. In the case that this fast test does not indicate convergence, a slight tightening of all parameters in a single additional calculation is advisable. If this is not feasible or does not confirm convergence, it is necessary to increase individual parameters in several different calculations to confirm convergence. When performing convergence testing with respect to individual parameters, we recommend halving time steps, increasing low-temperature corrections by one, and increasing rank or depth by two to five. The default parameters which control the behavior of the adaptive rank propagator in \texttt{TENSO} should be well-suited for most cases but can be adjusted as needed. Table \ref{tab:defaults} summarizes information about defaults and convergence checking for individual parameters.
\begin{table}[htb]
\caption{Recommendations for initial choice of important convergence settings as well as the suggested modification to make to the parameter during a convergence check.}
\label{tab:defaults}
\centering
\begin{tabularx}{\linewidth}{cXX}
\hline \hline
Parameter & Recommendation & To test convergence\\
\hline
dt & $1/20$ of the fastest oscillation period & halve\\
\hline
rank & 32 (fixed rank methods only) & increase by 2 to 5 as feasible\\
\hline
dim & 25 & increase by 2 to 5 as feasible\\
\hline
n\_ltc & 1 for high temperature, 3 for low & increase by 1\\
\hline
ode\_rtol & 1e-5 (default)& decrease by 1 order of magnitude\\
\hline
ode\_atol & 1e-7 (default)& decrease by 1 order of magnitude\\
\hline
max\_auxiliary\_rank & 32 (default) & increase by 5 to 10 as feasible\\
\hline
ps2\_atol & 1e-7 (default)& halve\\
\hline
ps2\_ratio & 2 (default)& adjust ps2\_atol instead\\
\hline \hline
\end{tabularx}
\end{table}

\section{Resources and Scaling}
\label{sec:scaling}
The resource requirements of \texttt{TENSO} depend on the system, convergence parameters, propagation strategy, and hardware. We recommend that \texttt{TENSO} run on CPUs. Whether available memory or computational time is the limiting factor for \texttt{TENSO} simulations will depend on the particular system, as the computation time required depends not only on the size of the tensor network but also on the relative ease of integrating the equation of motion. Stiff equations of motion require very small time steps leading to slow calculations. 

Parallelization in \texttt{TENSO} is handled by the pytorch backend, which implements parallel operations for some linear algebra tasks. By default, the backend in \texttt{TENSO} will begin parallel computations on as many cores as are available. We observe that this enhancement is typically capped at 2-4 threads. Some calculations with ps1 or ps2 propagation and very large ranks, on the order of 100, may benefit from additional threads. The number of threads can be controlled by adding \lstinline{_opt.set_num_threads(n)} where $n$ is the desired number, in \texttt{libs/backend}.

\begin{figure*}
    \centering
    \includegraphics[width=0.95\linewidth]{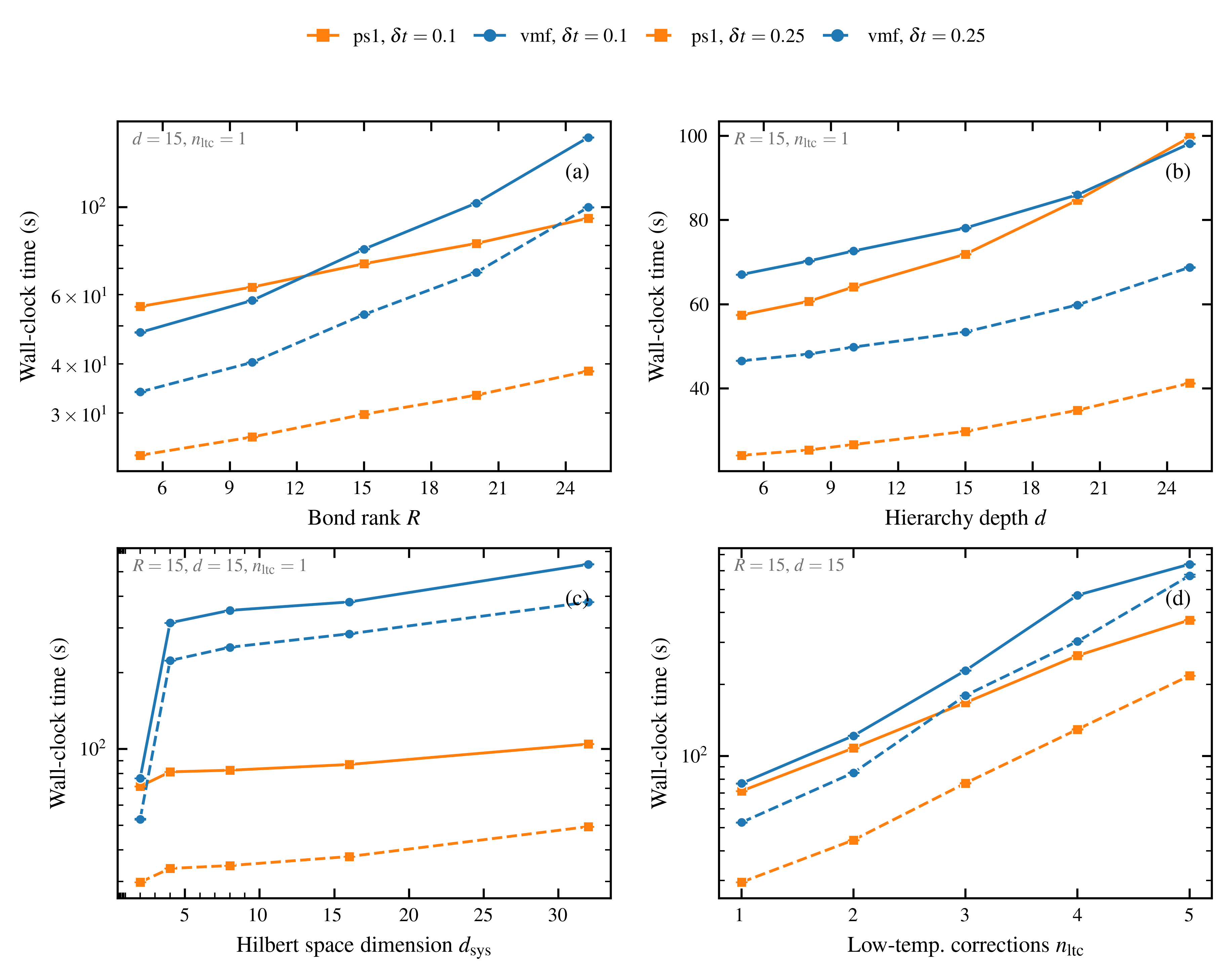}
    \caption{\textbf{Wall-clock time benchmarks comparing \lstinline{'simple'} propagation using the ps1 and vmf integrators at two time steps ($\delta t = 0.1$~fs and $\delta t = 0.25$~fs) as a function of four key simulation parameters.} \textbf{(a)} Scaling with bond rank $R$ at fixed hierarchy depth $d = 15$ and $n_{\text{ltc}} = 1$. \textbf{(b)} Scaling with hierarchy depth $d$ at fixed $R = 15$ and $n_{\text{ltc}} = 1$. \textbf{(c)} Scaling with system Hilbert space dimension $d_{\text{sys}}$ at fixed $R = 15$, $d = 15$, and $n_{\text{ltc}} = 1$. Dimension is increased by adding an additional, identical spin. All spins are coupled via $\sigma_{z}^{(i)}/2$ to a common bath, analogously to the qubits in Section \ref{sec:esd}, but not to each other. \textbf{(d)} Scaling with the number of low-temperature corrections $n_{\text{ltc}}$ at fixed $R = 15$ and $d = 15$.
The vmf integrator generally exhibits steeper scaling with increasing parameter values compared to ps1, while the larger time step $\delta t = 0.25$~fs consistently reduces runtime for both methods.
The ps1 integrator demonstrates substantially more favorable scaling with $d_{\text{sys}}$ (panel~c), remaining nearly flat up to $d_{\text{sys}} \approx 16$, whereas vmf shows a sharp increase beyond $d_{\text{sys}} = 2$.}
    \label{fig:scaling}
\end{figure*}

We now use the case of the spin-boson problem considered in Section \ref{sec:sbp1} to investigate the scaling of computational time. Many factors influence the computational requirements of \texttt{TENSO} and it is challenging to formulate simple rules, but some observations can be drawn. Figure \ref{fig:scaling} compares the wall time for computations with varying bond ranks, hierarchy depths, system sizes, and number of low temperature corrections for vmf and ps1 computations with identical time steps, either \lstinline{dt=0.1} or \lstinline{dt=0.25}, propagated for $100 \; \mathrm{fs}$ with a single thread on a Intel Xeon Gold 6448Y processor. All calculations are performed with the \lstinline{'simple'} propagation scheme with fixed ranks. Overall, straight vmf is usually slower than ps1 as it requires extensive regularization at the early stages of the dynamics. 
Further, the size of the initial time step can have significant impact on the efficiency of adaptive time step algorithms, in some cases changing whether vmf or ps1 is more efficient.

Increasing the number of low temperature corrections and adding additional bath features increases the wall time most dramatically. Increasing the system Hilbert space size by including additional, identical spin degrees of freedom to the system may also result in significant increases in wall time. In ps1 and vmf, computational cost increases with Hilbert space dimension $d$, as expected. For vmf, there is an additional computational cost increase with the first additional spin due to the introduction of a new timescale to the adaptive timestep method. By contrast, increasing rank and hierarchy depth leads to significantly more modest increases. The system bath coupling strength, characterized by the reorganization energy of the bath, also influences wall time (not shown). Wall time usually increases with coupling strength, but we have observed that the relationship can be non-monotonic, implying that certain weak coupling cases are more difficult to integrate than certain strong coupling cases.

\section{\texttt{TENSO} Structure}
\label{sec:software}

Having demonstrated applications of \texttt{TENSO} in a variety of prototypical quantum dynamics problems, we now discuss those details of the implementation of \texttt{TENSO} which are necessary to understand in order to modify the framework or adjust functionality. Documentation for TENSO is available at \url{https://ifgroup.github.io/pytenso/}. 

\begin{figure}[htb!]
    \centering
    \includegraphics[width=\linewidth]{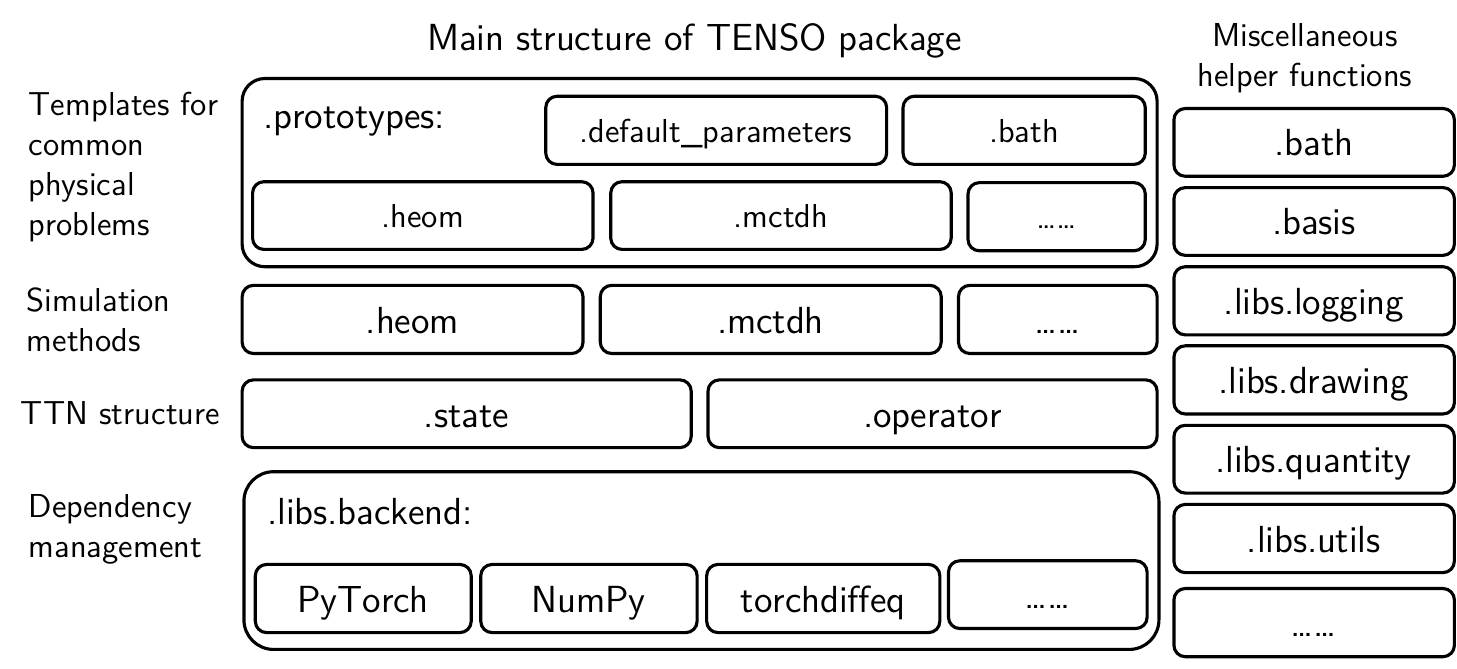}
    \caption{\textbf{Structure of \texttt{TENSO}.} 
    The code \texttt{TENSO} contains four main layers: (1) In \texttt{tenso.libs.backend} the needed data structures for handling a tensor array are imported from PyTorch and NumPy. 
    (2) The tree tensor network (TTN) structure layer defines the data structures needed for a TTN state and the sum-of-product (SoP) operator, as well as the decomposed master equation and propagation schemes based on the TTN state and SoP dynamics generator.
    (3) In the simulation method layer, the physical master equations are specified in the TTN decomposition and the SoP dynamics generator.
    (4) The templates in the \texttt{prototypes} subpackage are interfaces to easily use specific master equation methods.
   \texttt{TENSO}'s structure allows users to access high-level functions to quickly perform calculations and allows developers easy access to low-level internal structures to build extensions and modifications.
    }
    \label{fig:code}
\end{figure}

The implementation of TENSO contains four layers, shown in Fig. \ref{fig:code}. Layer 4 contains interfaces and templates and is the only layer with which most users will need to interact. It is designed to shield users from the more complex layers 1-3. Layer 3 includes implementations of HEOM and MCTDH and routines to set up the tensor network and the propagator which evolves the network in time. Layer 3 builds on the generic functions of layer 2, which provide a general framework for the evolution of any master equation with a sum-of-products form operator. Layer 2 requires libraries such as pytorch, and depends on the first layer to provide interfaces to these libraries. Some additional helper functions support multiple layers. We will discuss these layers in ascending order. Note that much of \texttt{TENSO} takes an object oriented approach and some familiarity with object oriented programming is assumed in this section.

\subsection{Layer 1: Interfaces with Backend Libraries}
In \texttt{tenso.backend}, dependencies and global package settings for pytorch, numpy, and torchdiffeq are handled. This includes specifying processor options for pytorch and importing or implementing integration methods. This file provides significant details on available integration methods other than the default adaptive time step Runge-Kutta.\cite{Dormand1980} All \texttt{TENSO} layers above rely on the basic linear algebra routines and pytorch settings in \texttt{tenso.backend} and a change to settings here is effective everywhere.

\subsection{Layer 2: TTN Decomposition}
The general master equations which \texttt{TENSO} is designed to address are given by Eq. \ref{eq:master_lvn} and Eq. \ref{eq:omega_tensor}.
The implementation of the efficient propagation for general master equations is based on two components, the TTN representation of the $\Omega(t)$ tensor and the SoP form of the generator of the dynamics, $\mathcal{L}(t)$. The second layer of \texttt{TENSO} implements the tensor network decomposition framework for a generic master equation with a generator in sum-of-products form. Layer 3 above depends on the \texttt{Node}, \texttt{End}, \texttt{Point} and \texttt{Frame} classes in layer 2 to describe the TTN structure and the associated classes for \texttt{Model} and \texttt{Propagator}.

In the \texttt{tenso.state} module, we implement the data structure for $\Omega(t)$. 
Specifically, \texttt{tenso.state.pureframe} defines the infrastructure needed to represent the underlying graph structure of the tensor network. The python class \texttt{Node} implements a node in the graph and each instantiated \texttt{Node} object is associated with a core tensor of the decomposition. The class \texttt{End} implements the open bonds in the tensor network. Both \texttt{Node} and \texttt{End} objects keep track of their neighbors through links, edges in the graph, and both are derived from the abstract \texttt{Point} class. A \texttt{Node} object may have more than one link, but an \texttt{End} object must have only one. The class \texttt{Frame} encodes the full graph of the topology of the tensor network. In \texttt{tenso.state.puremodel} we implement the tree tensor network vector $\Omega(t)$ in the class \texttt{Model}. A \texttt{Model} object includes a \texttt{Frame} object describing the TTN structure and data structures to associate a tensor valuation with every \texttt{Node} object in the \texttt{Frame}.

To specify the SoP generator of $\mathcal{L}(t)$ we implement the class \texttt{SparseSPO} in  \texttt{tenso.operator.sparse}. This class contains a list of dictionaries and each dictionary represents an operator product in the sum. The keys are the names of degrees of freedom to operate on and the values are the corresponding local operators represented by matrices.
We also implement the TDVP-based propagator class \texttt{SparsePropagator} in  \texttt{tenso.operator.sparse}.
This class generates the needed iterators for propagation with a given TTN vector of type \texttt{tenso.state.puremodel.Model} and a given SoP  generator of type \texttt{tenso.operator.sparse.SparsePropagator}.
Specific algorithms such as direct integration and projector-splitting algorithms\cite{chen2025} are implemented as methods in \texttt{SparsePropagator}.

\subsection{Layer 3: Implementing Physical Master Equations as a TTN Master Equation}
 Layer 3 of \texttt{TENSO} builds specific master equations on top of the generic classes provided by the second layer, providing rules to construct specific instances of the \texttt{tenso.operator.sparse.SparsePropagator} and \texttt{tenso.state.puremodel.Model} classes that are relevant to the desired master equations. \texttt{TENSO} currently implements HEOM in \texttt{tenso.heom} and ML-MCTDH in \texttt{tenso.mctdh}. In \texttt{tenso.heom.eom} we implement HEOM for a system coupled to one bath and in \texttt{tenso.heom.meom} we implement HEOM for a system coupled to multiple baths. 
Regardless of the master equation method, the class \texttt{Hierarchy} in the \texttt{meom} or \texttt{eom} module generates the needed data structure to construct a \texttt{tenso.operator.sparse.SparsePropagator}, as well as a \texttt{tenso.state.puremodel.Model}.
The class \texttt{FrameFactory} implements construction of the TTN topologies for $\Omega(t)$, including the balanced tensor tree and tensor train topologies. 

\subsection{Layer 4: Templates for Common Problems}
Layer 4 of \texttt{TENSO} provides a user-friendly interface, allowing users to avoid interacting with the more complex third layer while initializing a calculation. Interface templates for easy access to HEOM and MCTDH methods relevant to common physical problems are  implemented in the subpackage \texttt{tenso.prototypes}. These functions hide all details of the calculation set up to facilitate ease of use.
The \texttt{system\_multibath} function employed in the examples is located in \texttt{tenso.prototypes.heom}. This interface sets up the system state and propagator, including the tensor network topology requested.
Similarly, the helper function \texttt{gen\_bcf} employed to generate the composite spectral density is implemented in \texttt{tenso.prototypes.bath}. The default values of parameters are also found in this layer in \texttt{tenso.prototypes.default\_parameters}.

\subsection{Miscellaneous and Helper Functions}
Additional capabilities are provided by helper functions. These include algorithms for traversing over a tree to visit all of the nodes and related tasks in \texttt{tenso.libs.utils}, and options to use a discrete variable representation (DVR) basis, rather than the default number basis, in the master equations, in \texttt{tenso.basis}. The most important of these auxiliary functions are those which handle the correlation function calculations necessary for computation. The correlation function object \texttt{Correlation} is found in \texttt{tenso.bath.correlation}. The necessary decomposition of the Bose-Einstein distribution is found in \texttt{tenso.bath.distribution} and \texttt{tenso.bath.sd} handles the definitions of different spectral densities.

\section{Advanced Features}
\label{sec:advanced}

\texttt{TENSO} is an adaptable code. The preceding sections demonstrated its primary capabilities, features which are critical and available without modification of any part of \texttt{TENSO}. This section outlines how to access niche features and make simple modifications to the code to address specialized applications. We will demonstrate how to implement a custom topology for a TTN, how to modify the metric to adjust the variant of HEOM, and detail some more advanced options for the functions shown in the examples.

\subsection{Checkpoint Files}
It is possible to save a checkpoint file for a calculation. This checkpoint file will include all the information stored in the TTN. This behavior is requested by setting the parameter, \lstinline{save_checkpoint_to_file=True} when calling \lstinline{system_multibath}. Similarly, to load a checkpoint file, set \lstinline{load_checkpoint_from_file=True}. The checkpoint file's name will be the same as the output file but the extension will be '.pt.' Note that a calculation instructed to read from and write to a checkpoint file will first read and then overwrite the checkpoint. 

When loading from a checkpoint file, the initial state, Hamiltonian, operators, and bath correlation passed to \lstinline{system_multibath} will not be used for initialization but must be present and match those for the loaded checkpoint file. It is not advisable to change any parameters in the propagator except for the time step or end time, noting that propagation is always treated as beginning from time zero. The hierarchy depth and rank settings, in particular, must not be modified.

\subsection{Custom TTN Implementation}

\texttt{TENSO} enables propagation of TTN of arbitrary order and structure. It offers two predefined TTN structures, the TT and $n$-ary BTT requested by setting \lstinline{frame_method="train"} and \lstinline{frame_method="treen"} respectively, (i.e. \lstinline{frame_method="tree3"}). TENSO supports arbitrary TTN structures specified by defining a root, a list of $K$-nodes, and the links between them. A new TTN structure is defined in the \texttt{FrameFactory} class of the file defining the relevant equation of motion, meaning either MCTDH or HEOM (\texttt{src/tenso/heom/meom.py} or \texttt{src/tenso/mctdh/eom.py} for HEOM and MCTDH respectively) using the methods \lstinline{new_node()} and \lstinline{add_link()}.

\begin{figure}[htb!]
    \centering
    \includegraphics[width=0.75\linewidth]{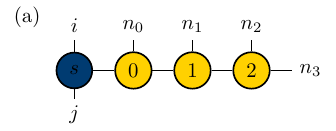} 
    \includegraphics[width=0.75\linewidth]{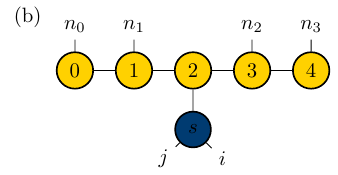}
    \caption{\textbf{TT Structures}: (a) Typical TT structure for a 4-feature bath. (b) Custom TT structure for a 4-feature bath. Nodes are represented by circles, system degrees of freedom are $i$ and $j$ and bath degrees of freedom are $n_i$ for $0 \le i \le 3.$}
    \label{fig:ttstructures}
\end{figure}

Listing~\ref{lst:customtn1} illustrates the implementation of a custom TT for a scenario with four total bath features in which the root node is placed in the center of the train rather than at either end.  Once the frame ``custom" is defined, the corresponding option must be registered in the frame selection logic for \lstinline{system_multibath} in \texttt{src/tenso/prototypes/heom.py} (or \texttt{src/tenso/prototypes/mctdh.py} for an MCTDH implementation) as shown in listing~\ref{lst:customtn2}. The user can request the custom tree by passing \lstinline{frame_method="custom",} to \lstinline{system_multibath}. The graphical representation of TENSO's default TT and the ``custom" TT is shown in Fig.~\ref{fig:ttstructures} where $n_i$ for $0 \le i \le 3$ are bath degrees of freedom and $i$ and $j$ are the system degrees of freedom.
\begin{lstlisting}[caption={\textbf{Custom TN Construction.} Python code defining a custom tensor network frame.}, label={lst:customtn1}]
# self refers to the FrameFactory instance
def custom(self) -> tuple[Frame, Node]:
    k_max = self.bath_dof
    frame = Frame()
    # Instantiate nodes for the custom train
    custom_train_nodes = [self._new_node() for _ in range(k_max + 2)]
    root = custom_train_nodes[-1]
    
    # Specifying the location of the leaves representing the system degrees of freedom
    frame.add_link(root, self.sys_ket_end)
    frame.add_link(root, self.sys_bra_end)
    # Adding links between neighboring nodes
    frame.add_link(custom_train_nodes[0], custom_train_nodes[1])
    frame.add_link(custom_train_nodes[1], custom_train_nodes[2])
    frame.add_link(custom_train_nodes[2], custom_train_nodes[3])
    frame.add_link(custom_train_nodes[3], custom_train_nodes[4])
    frame.add_link(custom_train_nodes[2], custom_train_nodes[-1])
    
    # Specifying the location of the leaves representing bath degrees of freedom
    frame.add_link(custom_train_nodes[0], self.chained_bath_ends[0])
    frame.add_link(custom_train_nodes[1], self.chained_bath_ends[1])
    frame.add_link(custom_train_nodes[3], self.chained_bath_ends[2])
    frame.add_link(custom_train_nodes[4], self.chained_bath_ends[3])
    
    return frame, root
\end{lstlisting}

\begin{lstlisting}[caption={\textbf{Method Integration.} Python code for registering the custom method.}, label={lst:customtn2}]
    # HEOM frame configuration:
    frame_method = parameters['frame_method']
    htd = FrameFactory(bath_correlation.k_max)
    
    if frame_method.lower() == 'train':
        frame, root = htd.train()
    elif frame_method.lower().startswith('tree'):
        n_ary = int(frame_method[4:])
        frame, root = htd.tree(n_ary=n_ary)
    elif frame_method.lower() == 'naive':
        frame, root = htd.naive()
    # Add these two lines
    elif frame_method.lower() == 'custom':
        frame, root = htd.custom()
    else:
        raise NotImplementedError(f'No frame_method {frame_method}.')
\end{lstlisting}

\subsection{Custom BCF Implementation}

The interface function \lstinline{gen_bcf} handles the details of decomposing a BCF into the needed sum of exponentials form when its spectral density consists of DL and Brownian terms. It is possible to add custom spectral densities to $\texttt{src/tenso/bath/sd.py}$. In \texttt{TENSO}, a spectral density is a class which must extend the abstract \lstinline{SpectralDensity} class. A custom spectral density class must implement a constructor, a function returning the spectral density at a given frequency $\omega$, and function returning a tuple containing lists of the residues and poles. Arguments accepting the new form of spectral density must be added to \texttt{src/tenso/prototypes/bath.py} in \lstinline{gen_bcf} along with an option to initialize objects of the custom spectral density type and add them to the list of spectral densities used to generate the BCF.  

A custom correlation function can be specified more easily if the complex coefficients, $c_k$ and $\gamma_k$ are known.  The helper function \lstinline{manual_corr_setup} in the \lstinline{Correlation} class in \texttt{tenso.bath.correlation} performs this function. The process necessary to perform the manual initialization is shown in Listing \ref{lst:customcorr}. The \lstinline{bath_simulation} produced can be passed to \lstinline{system_multibath} as usual. Note that in \texttt{TENSO} $c_k$ have units of energy squared and the $\gamma_k$ have units of energy. For real $\gamma_k$ only one feature will be introduced. For complex $\gamma_k$, the function will introduce a second feature with an exponent of $\gamma_k^*$. 
This may lead to redundant features when both $\gamma_k$ and $\gamma_k^\star$ are needed to define $C(t)$.
\begin{lstlisting}[caption={\textbf{Custom Correlation Function.} Code to initialize a custom correlation function.}, label={lst:customcorr}]
from tenso.bath.correlation import Correlation
bath_simulation = Correlation() # Initialize an empty correlation object
# Replace contents of these lists with all the complex coefficients of the custom correlation function
c_coefficients = [c_1, c_2, c_3] 
gamma_coefficients = [g_1, g_2, g_3]
# unit_convert=True indicates coefficients should be converted to internal TENSO units
bath_simulation.manual_corr_setup(c_coefficients,gamma_coefficients,unit_convert=True)
\end{lstlisting}

\subsection{Modifying the HEOM Variant}

The metric for the operator of the $k$th bexciton or bath feature is $\hat{z}_k|n_k\rangle=z_{k, n_k}|n_k\rangle$, where $|n_k\rangle$ is the excitation of the $k$th bexciton, $\hat{n}_k|n_k\rangle=n_k|n_k\rangle$. The commutator $[\hat{z}_k,\hat{n}_k]=0$ as these operators share a common eigenbasis. The dissipator $\mathcal{D}_k$ in Eq. \ref{eq:dissipator} associated with the $k$th bexciton exhibits a clear dependence on the metric.
While every metric choice formally leads to identical physical results, different metrics correspond to distinct HEOM variants and specific choices may generate numerical instabilities. In TENSO, alternative HEOM variants may be realized by implementing a custom metric in the file defining the relevant equation of motion. A new metric for single-bath HEOM is defined in \texttt{src/tenso/heom/eom.py}, as shown in Listing \ref{lst:custommetric}. The custom metric is selected with the argument \lstinline{metric='custom'} in the propagator.
\begin{lstlisting}[caption={\textbf{Custom Metric Implementation.} Python code for defining the metric factor.}, label={lst:custommetric}]
@staticmethod
def _adm_factor(k: int, c: Correlation, metric: Literal['re', 'abs'] | complex = 're'):
    c_k = c.coefficients[k]
    cc_k = c.conj_coefficents[k]
    
    if metric == 're':
        f_k = np.sqrt(abs(c_k.real + cc_k.real) / 2.0)
        if (c_k.imag + cc_k.imag) > EPSILON:
            f_k *= -1.0
    elif metric == 'abs':
        f_k = np.sqrt((abs(c_k) + abs(cc_k)) / 2.0)
    elif metric == 'custom':
        # Define custom metric here
        f_k = 1.0 # Placeholder
    else:
        f_k = complex(metric)
        
... # Remainder of the function is not modified
    return f_k
\end{lstlisting}

\subsection{Propagation Scheme and Troubleshooting}
\label{sec:prop_params}
We now review several important parameters governing the accuracy and efficiency of \texttt{TENSO} propagation, building on Section \ref{sec:defaults}.

The primary parameter for adjustment of \texttt{TENSO}'s propagation strategy is \lstinline{stepwise_method} in \lstinline{system_multibath} which determines the scheme for numerical propagation. By default, the \lstinline{stepwise_method="mix"} method is selected. This approach represents a hybrid scheme, typically combining a projector-splitting algorithm for the initial times followed by the direct integration method.\cite{chen2025} The other option is \lstinline{stepwise_method="simple"} in which case only one propagation method is used for the entire simulation, selected by setting the argument \lstinline{ps_method} equal to \lstinline{"vmf"} for variable-mean field direct propagation, \lstinline{"ps1"} for the static rank projector splitting method, and \lstinline{"ps2"} for the adaptive rank projector splitting method.

When \lstinline{"mix"} is selected, the initial phase of the propagation is controlled by the \lstinline{system_multibath} parameter \lstinline{auxiliary_ps_method}. This specifies the propagation algorithm. The default is the adaptive rank ps2 and the \lstinline{max_auxiliary_rank} parameter sets the upper bound on the tensor rank. When the rank limit is reached the propagation method changes to the method specified by the \lstinline{ps_method} parameter, with \lstinline{"vmf"} being the default. 

The accuracy of the non-adaptive rank projector splitting algorithm, ps1, is determined by the tensor rank. In a mixed propagation beginning with ps2 and transitioning to ps1 or vmf, the second method will use the final ranks selected by ps2. Otherwise, the value specified by the parameter \lstinline{rank} is used. For the rank-adaptive ps2 propagator, accuracy and efficiency are determined by the maximum size of singular values (SVs) truncated during the algorithm. The argument \lstinline{ps2_atol} defines the truncation threshold for singular values. To mitigate the loss of significant correlations during truncation, a fraction of discarded SVs are reincorporated, with this fraction determined by the \lstinline{ps2_ratio}. The direct integration vmf method relies on standard ordinary differential equation solvers, with the method determined by the \lstinline{ode_method} parameter. The default integration method is \lstinline{dopri5}, a fifth order Runge-Kutta method, with additional options detailed in \texttt{tenso.libs.backend}. The error tolerances of the integrator are controlled by absolute, \lstinline{ode_atol}, and relative, \lstinline{ode_rtol}, tolerances. 

The direct integration scheme also requires regularization of the equations of motion.  The relevant parameters control the  tolerance, \lstinline{vmf_atol},  method \lstinline{vmf_reg_method}, and procedure \lstinline{vmf_reg_type}. Regularization is essential for the time evolution of the non-root tensors which require the inversion of a matrix that may be ill-conditioned. The \lstinline{vmf_atol} parameter acts as a regularization term added to the diagonal elements of the ill-conditioned matrix to ensure invertibility. This regularization technique is always employed when the propagation scheme is set to vmf. Note that seriously ill-conditioned matrices are most likely to arise at the start of the simulation which is why the default mixed propagation scheme does not start with the vmf method.

The reliability and accuracy of \texttt{TENSO}'s tensor network propagation strategies depends on interactions between control parameters and on the structure of the tensor network. TTN-HEOM equations of motion and their implementation into \texttt{TENSO} admit arbitrary tensor tree structures with tensors of any order. In practice, however, some choices may outperform others. There is not yet means to predict which tree structures may be more favorable. \texttt{TENSO}’s default balanced binary tree combined with mixed ps2 to vmf propagation is the recommended initial treatment of a system. In the case that a simulation is unsuccessful, it is possible that the problem is unsuitable for a tensor network compression, or it is possible that different control parameters are required for stable propagation. Decreasing the time step, changing between vmf, ps1 or ps2 propagation, or selecting an alternative integrator from \texttt{src/tenso/libs/backend} may improve the situation. Such cases should be carefully scrutinized for convergence.

\subsection{Modifying Default Units}
\texttt{TENSO}'s default units are defined in \texttt{prototypes/default\_parameters.py}. The units which \texttt{TENSO} assumes when reading input from \lstinline{system_multibath} and \lstinline{bath_correlation} are defined in a dictionary, \lstinline{default_units}, where keys are properties and values are strings corresponding to physical constants imported from SciPy, with the available list detailed in \texttt{libs/quantity.py}. Options for energy units include the default inverse centimeters, \lstinline{"/cm"}, joules, \lstinline{"J"}, electron volts, \lstinline{"eV"}, and millielectron volts, \lstinline{"meV"}.  Options for time include the default femtoseconds, \lstinline{"fs"}, seconds \lstinline{"s"}, and picoseconds \lstinline{"ps"}. Inverse temperature, not temperature, is defined, with the only option being the default, inverse Kelvin, \lstinline{"/K"}.

\texttt{TENSO} converts input into internal units and then scales energy and time by the factor associated with \lstinline{"unital_energy"} in the \lstinline{default_units} dictionary. Specifically any units of time or inverse temperature will be multiplied by this factor and any units of energy will be divided by this factor. The default value is 1000. Adjusting this value may alleviate numerical problems when dealing with energies which are very large or very small in absolute terms.

\section{Conclusion}

By combining the hierarchical equations of motion with tensor network representations \texttt{TENSO} provides an efficient and systematically improvable framework that alleviates the curse of dimensionality inherent in conventional HEOM simulations. The resulting approach retains the formal exactness of HEOM while extending its applicability to complex spectral densities and arbitrary Hamiltonians, including those with explicit time dependence.

We demonstrated the versatility of \texttt{TENSO} through a series of representative examples ranging from structured spin--boson models and driven two-level systems to the Fenna--Matthews--Olson complex to scenarios exhibiting entanglement sudden death. In each case, \texttt{TENSO} was able to capture rich dynamical behavior with reduced computational overhead, highlighting the efficiency of the TTN-based propagation scheme.

A key strength of \texttt{TENSO} lies in its modular and extensible design providing both accessibility for non-specialists and flexibility for advanced users developing new approaches. By making the underlying tensor operations and graph-based decomposition structures available for advanced users, \texttt{TENSO} serves not only as a powerful simulation tool but also as a platform for further method development. It can easily be adapted to employ specialized and efficient correlation function decomposition schemes such as ESPRIT\cite{roy1990,potts2013,takahashi2024_high_accuracy} and A4\cite{hunt2026} as needed for instance to address low temperature cases difficult for traditional decomposition schemes. \texttt{TENSO}'s cost saving strategy for HEOM is compatible with alternative approaches to reduce the computational cost of capturing structured baths, including pseudomode,\cite{garraway1997,park2024,lorenzoni2025,huan2026} two-particle approximation methods\cite{song2015,bhattacharyya2024}, and other strategies. The combination of \texttt{TENSO} with these additional strategies offers fertile ground for the development of increasingly powerful computational strategies for HEOM-based simulations. Future studies will investigate the limits of the tensor network decomposition of HEOM.

Taken together, its features position \texttt{TENSO} advantageously as a general and efficient framework for simulating quantum dynamics in complex environments.

\section*{Data Availability}
TENSO is available under the MIT license. The python code for TENSO, including examples, is available at \url{https://github.com/ifgroup/pytenso.} The data underlying this study are available in the published article.

\section*{Acknowledgments}
  This material is based on work supported by the U.S.\ Department of Energy, Office of Science, Office of Basic Energy Sciences, Quantum Information Science Research in Chemical Sciences, Geosciences, and Biosciences Program under Award No.~DE-SC0025334.

\bibliography{tensorefs}

\end{document}